\RequirePackage{lineno}
\documentclass[prd,twocolumn,floatfix,amsmath,amssymb,floatfix,nofootinbib]{revtex4}
\usepackage{graphicx,color,dcolumn,booktabs,bm}
\usepackage{longtable,lscape,comment,braket}
\usepackage{txfonts}
\usepackage{overpic}
\usepackage{amssymb}
\usepackage{indentfirst}
\usepackage{feynmf}   %{feynmp}
\usepackage{slashed}  %for Feynman symbols
\usepackage{cases}
\usepackage{epstopdf}
\usepackage{psfrag}
\usepackage{subfigure}
\usepackage{threeparttable}
\usepackage{color}
\usepackage{gensymb}
\usepackage{multirow}
\usepackage{graphicx,color,dcolumn,booktabs,bm}
\usepackage[title]{appendix}
\def\lrpartial{\buildrel\leftrightarrow\over\partial}
\graphicspath{{Figures/}} %
\usepackage[colorlinks, citecolor=blue,anchorcolor=red,menucolor=red, linkcolor=red,filecolor=red,runcolor=red,urlcolor=blue,frenchlinks=red]{hyperref}

\usepackage{ulem}

\makeatletter
\renewcommand{\maketag@@@}[1]{\hbox{\m@th\normalsize\normalfont#1}}%
\makeatother
\def \ff {\phi\phi}

\def \ee {e^+e^-}
\def \kk {K^+K^-}

\def \kk {K^+K^-}

\def \ee {e^+e^-}

\newcommand{\chicj}{\chi_{cJ}}
\newcommand{\chiz}{\chi_{c0}}
\newcommand{\chio}{\chi_{c1}}
\newcommand{\chit}{\chi_{c2}}

\newcommand{\psip}{\psi(2S)}

\begin{document}
%\linenumbers
%\begin{CJK}{GBK}{}

\title{Detecting the polarization in $\chicj\to\ff $ decays to probe hadronic loop effect}
\author{Qi Huang$^{1}$}\email{huangqi@ucas.ac.cn}
\author{Jun-Zhang Wang$^{2,3}$}\email{wangjzh2012@lzu.edu.cn}
\author{Rong-Gang Ping$^{1,4}$}\email{pingrg@ihep.ac.cn}
\author{Xiang Liu$^{2,3,5}$}\email{xiangliu@lzu.edu.cn}
\affiliation{$^1$University of Chinese Academy of Sciences (UCAS), Beijing 100049, China\\
$^2$School of Physical Science and Technology, Lanzhou University, Lanzhou 730000, China\\
$^3$Research Center for Hadron and CSR Physics, Lanzhou University and Institute of Modern Physics of CAS, Lanzhou 730000, China\\
$^4$Institute of High Energy Physics, Chinese Academy of Sciences, P.O. Box 918(1), Beijing 100049, China\\
$^5$Lanzhou Center for Theoretical Physics, Key Laboratory of Theoretical Physics of Gansu Province, and Frontiers Science Center for Rare Isotopes, Lanzhou University, Lanzhou 730000, China}

\begin{abstract}
In this work, we show that detecting the polarization information of $\chi_{cJ}\to \phi\phi~(J=0,1,2)$ could be a practical way to test the hadronic loop effect in these decays. Our results show that the predicted ratios of helicity amplitudes are less dependent on the parameters involved in the calculation. The ratios are determined to be $|F_{1,1}^{(0)}/F_{0,0}^{(0)}| \approx 0.359$, $|F_{1,0}^{(1)}/F_{0,1}^{(1)}| =1$, $|F_{1,1}^{(1)}/F_{0,1}^{(1)}| =0$, $|F^{(2)}_{1,0}|/|F^{(2)}_{0,0}|=|F^{(2)}_{0,1}|/|F^{(2)}_{0,0}| \approx 1.285,~|F^{(2)}_{1,-1}|/|F^{(2)}_{0,0}|=|F^{(2)}_{-1,1}|/|F^{(2)}_{0,0}| \approx 5.110$ and $|F^{(2)}_{-1,-1}|/|F^{(2)}_{0,0}|=|F^{(2)}_{1,1}|/|F^{(2)}_{0,0}| \approx 0.465$.
By adopting these predicted ratios, we use the Monte-Carlo events to show that the moments $\langle t_{ij} \rangle$ can be used as observables to reveal the polarization transfer in these decays, and their distributions are directly related to the determination of helicity amplitudes. We suggest experiments like at the BESIII and the Belle II  performing a polarization analysis on the $\chi_{cJ}\to \phi\phi$ decay in the future, and the results are important to understand the decay mechanism underlying the $\chi_{cJ}$ decays.
\end{abstract}

%pacs{ 13.20.Gd, 13.30.Eg, 13.88.+e, 14.40.Pq }

\maketitle

\section{Introduction}\label{sec1}
How to quantitatively depict nonperturbative behavior of strong interaction at the medium and low energy region is a great challenge for us to understand the hadronic confinement and the related hadron properties. Though the foundation of lattice QCD has well established and commonly accepted as a reliable way to deal with these issues, its development is relatively slow compared to the accumulations of experimental data in last two decades \cite{ref1,ref2,ref3,ref4,ref5}. In recent years, more and more exotic hadronic states and exotic decays are reported, which enriches our knowledge about the hadron spectroscopy \cite{ref6,ref7,ref8,ref9,ref10}. To reveal the mechanisms underlying these new structures and their decay properties, the effective theory, QCD sum rule and other phenomenological methods are indispensable for bridging the strong interactions to the hadron nonpertubative properties.

Since the discovery of $J/\psi$ particle in 1974, the charm physics has been continuing to be the laboratory to study the properties of light hadron, charmed meson and charmonium. In this work, we suggest the experiments to measure the polarization transfer in the $\psi(2S)\to\gamma\chi_{cJ},~\chi_{cJ}\to\ff$ to figure out the decay mechanism of $\chi_{cJ}$.
Experimentally, the $\phi$ mesons can be reconstructed with the Kaon charged tracks registered in the detector. The narrow width of the $\chicj$ and $\phi$ mesons provide a beneficial selection criteria to reconstruct the decay chain. The branching fractions for the charmonium transition $\psip\to\gamma\chicj(J=0,1,2)$ are summed up to about 30\%, hence this transition could be the $\chicj$ factory, which can produce a large size of $\chicj$ events. For example, with accumulation of 3 billion $\psip$ events in the future \cite{Ablikim:2019hff}, it will produce 0.9 billion $\chicj$ events. The large size of $\chicj$ sample is essential to analyze the polarization transfer, since the experiment is dependent on the observation of polarization moments, which are constructed with the helicity angles in the cascade decays.

The $\chio\to\ff$ decay was first reported by the BESIII Collaboration in 2011 \cite{Ablikim:2011aa}. With observed 254 events in the $\ff$ mass spectrum, its branching fraction was significantly measured with the same order as that for the $\chiz\to\ff$ decay \cite{Ablikim:2011aa,Zyla:2020zbs}. This puts a great challenge for the theoretical investigation on its decay mechanism in the pQCD scenario. Since the strong decay conserves the parity, there is a strong constraint on the helicity selection rule for the $\chio\to\phi(\lambda_1)\phi(\lambda_2)$ decay, here $\lambda_1,\lambda_2$ denote the helicity values of $\phi$ mesons. Besides, there exists a further constraint, which is due to the requirement on the identical particle symmetry of $\ff$ system. These constraints result in the helicity amplitude vanishing with helicity selection rule $\lambda_1=\lambda_2$. The nonvanishing helicity amplitudes are those with $(\lambda_1,\lambda_2)=(\pm1,0)$ or $(0,\pm1)$. However, these helicity configurations violate the helicity selection rule in the $\chio$ decay. Their contributions to the partial decay width is suppressed by a factor of $(\Lambda_{\textrm{QCD}}/m_c)^{6}$, here $m_c$ is the mass of charm quark, and $\Lambda_{\textrm{QCD}}$ is the QCD energy scale. This strong suppression can be simply understood according to the Landau-Yang theorem \cite{Yang:1950rg}.

The same situation happens in the $\chit\to\phi(\lambda_1)\phi(\lambda_2)$ decay \cite{Ablikim:2018ogu}. The parity conservation and identical particle symmetry allow the existence of the nonvanishing amplitude for all helicity configurations with $|\lambda_1-\lambda_2|\le2$. But the helicity selection rule results in a suppression factor to the branching fraction according to the pQCD asymptotic behavior \cite{Chernyak:1981zz}, i.e.,   
\begin{eqnarray}
  \mathcal{B}(\chit \to \phi(\lambda_1)\phi(\lambda_2)) \sim \left(\frac{\Lambda_{\mathrm{QCD}}^2}{m_c^2}\right)^{|\lambda_1+\lambda_2|+2}.
\end{eqnarray}
The pQCD calculation determined its branching fraction to be $\mathcal{B}(\chi_{c2} \to \phi \phi) = 7.8 \times 10^{-4}$ \cite{Zhou:2004mw}, which is less than the average of measured ones, $(1.06 \pm 0.09) \times 10^{-3}$ \cite{Zyla:2020zbs}, by about 30\%.

To decode the nonperturbative mechanism underlying these decays, the hadronic loop mechanism (HLM) \cite{Liu:2006dq,Meissner:2010zz,Guo:2010ak,Liu:2009vv,Chen:2009ah,Guo:2010zk,Liu:2010um,Colangelo:2002mj,Chen:2011zv,Meng:2007tk,Meng:2008dd,Chen:2011qx,Chen:2014ccr,Chen:2011pv,Meng:2008bq,Liu:2009dr,Li:2013zcr,Cheng:2004ru,Simonov:2008qy,Chen:2010re,Li:2007au,Li:2011ssa} was introduced in Refs. \cite{Liu:2009vv,Chen:2009ah}. In this scenario, the $\ff$ production is supposed to be from the rescattering of two virtual charmed mesons by exchanging another charmed meson. It shows that under HLM the measured branching ratios can be well reproduced within experimental uncertainty. To some extent one can expect that the hadronic loop mechanism can model the long distance effects in the $\chicj$ decays. However, to comprehensively test the hadronic loop mechanism, it is desirable to give more predictions on the $\chicj$ decays besides the branching fraction.

In this work, we show that detecting the polarization information of $\chi_{cJ}\to \phi\phi$ can be an effective way to probe hadronic loop mechanism. Especially in the $\chit$ decays, the two $\phi$ decays provide us with rich spin observables. In the HLM scenario, these spin observables can be predicted by calculating the helicity amplitudes. The suppression of helicity selection rule is characterized by the ratios between the amplitudes of the $\phi$ longitudinal polarization components. Thus, experimental measurements on the ratios of helicity amplitudes can provide us with a very good platform to test if there exists the long distance contributions to the $\chi_{cJ}\to\phi\phi$ decays. With this motivation, we present a polarization analysis on the $\chi_{cJ}\to \phi\phi$ decays in the HLM scenario. We find that the obtained ratios of helicity amplitudes is less dependent on the parameters, i.e., their values change little with variations of the free parameters in the model. An ensemble of Monte-Carlo (MC) events are generated based on the amplitude ratios, and we show that some moments $\langle t_{ij}\rangle$ can be used as polarization observable. With this investigation, we strongly suggest the BESIII and Belle II to perform a measurement on the polarization in the $\chi_{cJ}\to \phi\phi$ decays, which may provide crucial test to the hadronic loop mechanism.

This paper is organized as follows. After the introduction, we present a polarization analysis of the decay $\chicj\to \ff\to2(\kk)$ in Sec. \ref{sec:polar}. Then, the calculation of $\chi_{cJ} \to \phi \phi$ in the charmed meson loop scenario is presented in Sec. \ref{sec:loop}. The numerical results are given in Sec. \ref{sec:nr}. Finally, this paper ends with a summary.

\section{Polarization analysis}\label{sec:polar}
We analyze the $\ff$ polarization started with $\ee$ beams. In the unpolarized $\ee$ collider, the production of $\psip$ particle is tensor polarized without longitudinal polarization \cite{pingrg}. Thus, the subsequent $\psip\to\gamma\chicj$ decay may transfer some polarization to the $\chicj$ states, which is manifested in the $\chicj\to\ff$ decay, showing up with a unflat angular distribution of the decayed $\phi$ meson.

To get the polarization or alignment information of $\ff$ system, one has to investigate its spin density matrix (SDM), which encodes the full polarization information transferred from the $\chicj$ decays. In experiment, the measurement on the $\ff$ SDM plays the role to study the $\chicj$ decay mechanism, given that the polarization patten is predicted based on the decay-dynamical models. We follow the standard way to construct the SDM for the identical particle $\ff$ system.

\begin{figure*}[htbp]
\centering
\mbox{\begin{overpic}[scale=0.8]{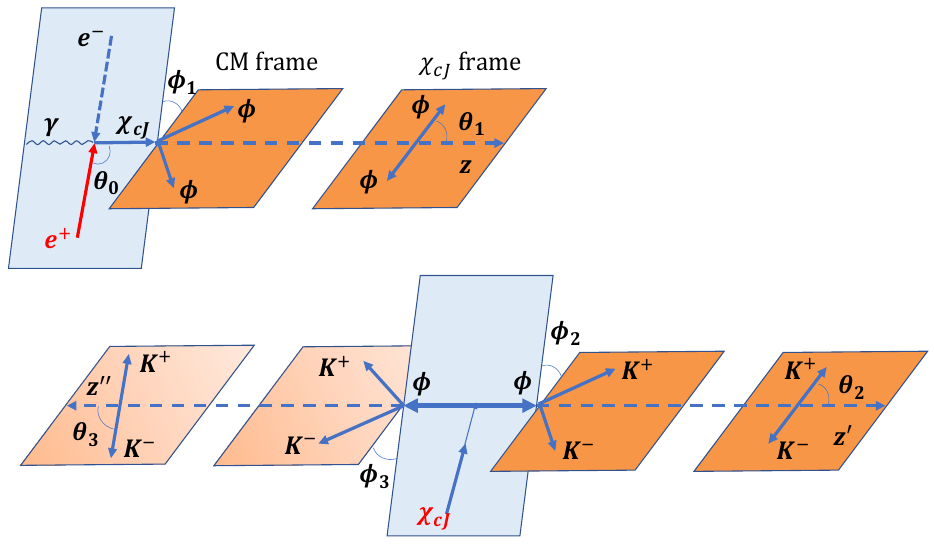} \end{overpic} }
{\caption{Helicity system and angles definition for the $\psip\to\gamma\chicj, \chicj\to\ff,\phi\to \kk$ decay.}\label{helsys}}
\end{figure*}

For a spin-$s$ particle, its spin density matrix is given in terms of multipole parameters, $r^L_M$, as \cite{doncel}
\begin{equation}
\rho = {1\over 2s+1}\left(\mathcal{I}+2s\sum_{L=1}^{2s}\sum_{M=-L}^{L}r^L_M Q[s,L,M]\right),
\end{equation}
where $\mathcal{I}$ denotes a $(2s+1)\times(2s+1)$ dimensional unit matrix, $L$ is the orbital angular momentum with its magnetic quantum number $M$ defined in the spherical tensor operator.
The SDM for $\ff$ system can be constructed from the $\phi$ individual ones, and an easy way is to decompose it into $Q$ matrices multiplied by a set of real parameters, which reads as
\begin{eqnarray}
\rho_{\ff}&=&\rho_{\phi}\otimes\rho_{\phi}\nonumber\\
&=&{1\over 9}\left[C_{00}\mathcal{I}_3\otimes \mathcal{I}_3+2\sum_{i=1}^8(C_{i,0} Q_i\otimes \mathcal{I}_3 \right.\nonumber\\
&&+\left. C_{0i}\mathcal{I}_3\otimes Q_i)+ 4\sum_{i,j=1}^8 C_{ij}Q_i\otimes Q_j \right].
\end{eqnarray}
Here, $\mathcal{I}_3$ denotes a $3\times3$ identity matrix. And the real parameters $C_{ij}$ is determined from the $\ff$ production process, which carry polarization information for the two $\phi$ mesons. $C_{i0}$ or $C_{0i}$ means that the polarization is detected only for one $\phi$ meson, while $C_{ij}$ measures the polarization correlation between two $\ff$ mesons. For $Q_i$, we define $Q_1=Q[1,1,-1],~Q_2=Q[1,1,0],~Q_3=Q[1,1,1],~Q_4=Q[1,2,-2],~Q_5=Q[1,2,-1],~Q_6=Q[1,2,0],~Q_7=Q[1,2,1],~Q_8=Q[1,2,2]$.

The polarization of $\ff$ system is unaccessible in a general purpose of electromagnetic spectrometer at the modern $\ee$ colliders. Nonetheless, the subsequential decay, $\phi\to \kk$, can be used as the polarimeter to measure the $\phi$ polarization by studying the implications of the decayed Kaon angular distribution.

We formulate the $\ff\to2(\kk)$ decays with helicity amplitude method, which is defined in the helicity system as shown in Fig. \ref{helsys}. One $\phi$ decaying into $\kk$ pair is described with helicity angles ($\theta_2,\phi_2$), where $\theta_2$ is the angle spanned between the directions of $K^+$ and the $\phi$ momenta, which are defined in the rest frames of their respective mother particles. The azimuthal angle $\phi_2$ is defined as the angle between the $\ff$ production plane and the $\phi$ decay plane. The helicity angles, $(\theta_3,\phi_3)$, describing another $\phi$ meson decay, is defined by the same rule (see Table \ref{angdef}). Then the joint angular distribution for $\ff\to2(\kk)$ reads as

\renewcommand\tabcolsep{0.13cm}
\renewcommand{\arraystretch}{1.50}
\begin{table}[!htpb]
\caption{Definitions of helicity angles and amplitudes in the $\chicj\to\ff$, and $\ff\to2(\kk)$ decays. }\label{angdef}
\begin{tabular}{ccc}
\hline\hline
Decay & Angles & Amplitude \\\hline
$\chi_{cJ}\to\phi(\lambda_1)\phi(\lambda_2)$ & $(\theta_1,\phi_1)$ & $F^{(J)}_{\lambda_1,\lambda_2}$ \\
$\phi(\lambda_1)\to\kk$ & $(\theta_2,\phi_2)$ & $f$ \\
$\phi(\lambda_2)\to\kk$ & $(\theta_3,\phi_3)$ & $f$ \\\hline\hline
\end{tabular}
\end{table}

\begin{eqnarray}
|\mathcal{M}|^2&\propto& \text{Tr}[\rho_{\ff}\cdot M_a\otimes M_b^\dag]\nonumber\\
&&=t_{00}C_{00} + \sum_{i=1}^8(t_{i0} C_{i0}+t_{0i} C_{0i}) + \sum_{i,j=1}^8 t_{ij}C_{ij}%\nonumber
\end{eqnarray}
with
\begin{eqnarray}
(M_a)_{\lambda_1,\lambda'_1}&=&D^{1*}_{\lambda_1,0}(\phi_2,\theta_2,0)D^{1}_{\lambda_1',0}(\phi_2,\theta_2,0)f^2,\\
(M_b)_{\lambda_2,\lambda'_2}&=&D^{1*}_{\lambda_2,0}(\phi_3,\theta_3,0)D^{1}_{\lambda_2',0}(\phi_3,\theta_3,0)f^2.
\end{eqnarray}
Since the helicity value $f$ is independent on the Kaon angles, it contributes a trivial constant factor to the angular distribution. Thus, for simplicity, we can take $f^2=1$. The joint angular distribution can be further decomposed into the $\ff$ polarization in terms of the real multipole parameters $C_{ij}$. The $t_{ij}$ factors play the role of the spin observables corresponding to the parameters $C_{ij}$. The term $t_{00}$ is the unpolarization cross section, while $t_{0L}(t_{L0})$ corresponds to the observable for detecting one $\phi$ polarization with rank $L$, and leaving another $\phi$ polarization being undetected. The term $t_{ij}$ denotes the spin correlation between the two $\phi$'s. Expressions of $t_{ij}$ factors are given in terms of angles $\theta_i$ and $\phi_i\,(i=2,3)$ as shown in Appendix \ref{appA}.

The multipole parameters, $C_{ij}$, in the $\rho_{\ff}$ SDM contain the dynamical information of the $\chicj\to\ff$ decays, which can be related to the helicity amplitudes $F^{(J)}_{\lambda_1,\lambda_2}$. Thus, any theoretical prediction on their values can be tested by measuring their spin observables in experiment.

We relate the parameter $C_{ij}$ to the helicity amplitude $F^{(J)}_{\lambda_1,\lambda_2}$ by calculating the spin density matrix $\rho_{\ff}$ of the decay $\chi_{cJ}\to\ff$, which reads as
\begin{equation}
\rho_{\ff} = N\cdot \rho_J\cdot N^\dag,
\end{equation}
where $\rho_J$ is a spin density matrix for $\chicj$ with $J=0,1,2$ for $\chiz$, $\chi_{c1}$ and $\chi_{c2}$, respectively. $N$ denotes decay matrix, which can be written as
\begin{eqnarray}
(N)_{\lambda_1\lambda_2\lambda'_1\lambda'_2,M}&=&D^{J*}_{M,\lambda_1-\lambda_2}(\phi_1,\theta_1,0)\\
&&\times D^{J}_{M,\lambda'_1-\lambda'_2}(\phi_1,\theta_1,0)F^{(J)\ast}_{\lambda_1,\lambda_2}F^{(J)}_{\lambda'_1,\lambda'_2},\nonumber
\end{eqnarray}
where $(\theta_1,\phi_1)$ are the helicity angles describing the $\phi$ meson flying direction as shown in Fig. \ref{helsys}. Azimuthal $\phi_1$ is defined as the angle between the $\phi$ production and decay planes, while $\theta_1$ is the angle spanned between the $\phi$ and $\chicj$ momenta. $F^{(J)}_{\lambda_1,\lambda_2}$ denotes the helicity amplitude in terms of two $\phi$ helicity values $\lambda_1$ and $\lambda_2$.

A special decay is $\chiz\to\ff$, where the spin density matrix of $\chiz$ is reduced to Kronecker delta function, {\it i.e.}, $\rho_{0}=\delta_{\lambda_1,\lambda_2}\delta_{\lambda'_1,\lambda'_2}$. Then the multipole parametes $C_{ij}$ are calculated to be
\begin{eqnarray}
  C_{00}&=&9\left|F^{(0)}_{0,0}\right|{}^2+18 \left|F^{(0)}_{1,1}\right|{}^2,\\
  C_{4 4}&=& -\frac{3}{2} \left|F^{(0)}_{1,1}\right|{}^2,\\
  C_{5 5}&=& -C_{77}=-\frac{3}{4} \left(F^{(0)\ast}_{1,1} F^{(0)}_{0,0}+F^{(0)\ast}_{0,0} F^{(0)}_{1,1}\right),\\
  C_{6 0}&=&C_{0 6}=\left|F^{(0)}_{1,1}\right|{}^2-\left|F^{(0)}_{0,0}\right|{}^2,\\
  C_{6 6}&=& \left|F^{(0)}_{0,0}\right|{}^2+\frac{\left|F^{(0)}_{1,1}\right|{}^2}{2},\\
  C_{8 8}&=& \frac{3 \left|F^{(0)}_{1,1}\right|{}^2}{2},
\end{eqnarray}
while other $C_{i,j}$ parameters are vanishing due to the spin-parity conservation in the $\chiz\to\ff$ decays.

Then, with the helicity amplitude $F^{(0)}_{\lambda_1,\lambda_2}$, the $\phi$ angular distribution from the $\chiz\to\ff$ decay can be expressed  as
\begin{eqnarray}
\mathcal{W}_0&\propto&\cos \left(\phi _{23}\right) \left[4 \sin ^2\left(\theta _2\right) \sin ^2\left(\theta _3\right) \cos
   \left(\phi _{23}\right) \left|F^{(0)}_{1,1}\right|{}^2\right.\nonumber\\
   &&+\left.\sin \left(2 \theta _2\right) \sin \left(2 \theta
   _3\right) 2\text{Re}(\text{F}^{(0)\ast}_{1,1} F^{(0)}_{0,0})\right]\\
   &&+4 \cos ^2\left(\theta _2\right)
   \cos ^2\left(\theta _3\right) \left|F^{(0)}_{0,0}\right|{}^2\nonumber,
\end{eqnarray}
where $\phi_{23}=\phi _2+\phi _3$.

One can see that the $\phi$ angular distribution for the $\chiz\to\ff$ decay is reduced to a uniform distribution either on the $\cos\theta_2(\cos\theta_3)$ or $\phi_2(\phi_3)$ observables alone. Spin correlation for $\ff$ system can only be observed by measuring a moment formed by the angles $\theta_i$ and $\phi_i\,(i=2,3)$ simultaneously.

The strong decay $\chi_{c1}\to\ff$ conserves the parity. Thus, the helicity amplitudes satisfy the relation $F^{(1)}_{-\lambda_1,-\lambda_2}=-F^{(1)}_{\lambda_1,\lambda_2}$, especially $F^{(1)}_{0,0}=0$. Then the amplitudes are reduced to three independent components, i.e., $F^{(1)}_{1,1},F^{(1)}_{1,0}$ and $F^{(1)}_{0,1}$. The matrix of helicity amplitudes is taken as
\begin{equation}
\left(F^{(1)}_{\lambda_1,\lambda_2}\right)=\left(
\begin{array}{ccc}
F^{(1)}_{1,1} & F^{(1)}_{1,0} &0\\
-F^{(1)}_{1,0}&0&-F^{(1)}_{0,1}\\
0&-F^{(1)}_{1,0}&-F^{(1)}_{1,1}.
\end{array}
\right).
\end{equation}
As for the $\chio$ production from the decay $\psip\to\gamma\chio$, its SDM is well defined and taken as $\rho_1={1\over 4}\text{diag}\{1,2,1\}$ \cite{pingrg} in its rest frame. Here, the nonvanishing parameters $C_{i0},~C_{0i}$ and $C_{ij}$ are calculated and given in Appendix \ref{appB}.

The $\phi$ meson has nonzero decay width, the masses of two $\ff$ may have different values from the $\chio$ decay in a given event. However, its narrow decay width allows us to treat the $\ff$  as an identical particle system statistically. Then, Exchanging two $\phi$ mesons yields asymmetry relation $F^{(1)}_{1,0}=-F^{(1)}_{0,1}$, and $F^{(1)}_{1,1}=0$, where the joint angular distribution is independent on the amplitude, and it reads
\begin{eqnarray}
\mathcal{W}_1&\propto& (2+\sin^2\theta_1)[\cos^2\theta_2\sin^2\theta_3+\sin^2\theta_2\cos^2\theta_3].\nonumber\\
\end{eqnarray}

Similarly, we perform the same analysis on the $\chi_{c2}\to\ff$ decay, and we take the $\chit$ SDM as $\rho_2={3\over 20}\text{diag}\{2,1,2/3,1,2\}$ \cite{pingrg}. Considering the parity conservation in this decay, one has the relation $F^{(2)}_{-\lambda_1,-\lambda_2}=F^{(2)}_{\lambda_1,\lambda_2}$, then the amplitude matrix is reduced to be
\begin{equation}
\left(F^{(2)}_{\lambda_1,\lambda_2}\right)=\left(
\begin{array}{ccc}
F^{(2)}_{ 1,1} & F^{(2)}_{ 1,0} & F^{(2)}_{ 1,-1}\\
F^{(2)}_{ 0,1} & F^{(2)}_{ 0,0} & F^{(2)}_{ 0,1}\\
F^{(2)}_{1,-1} & F^{(2)}_{ 1,0} & F^{(2)}_{ 1,1}
\end{array}
\right).
\end{equation}
With these considerations, the multipole parameters are calculated and given in Appendix \ref{appC}, and these expressions can be further simplified using the relation $F^{(2)}_{\lambda_1,\lambda_2}=F^{(2)}_{\lambda_2,\lambda_1}$ if one takes the $\ff$ as an identical particle system.

\section{Meson loop  effects in $\chicj\to\ff$ decay}\label{sec:loop}

\begin{center}
\begin{figure}[htbp]
\scalebox{0.075}{\includegraphics{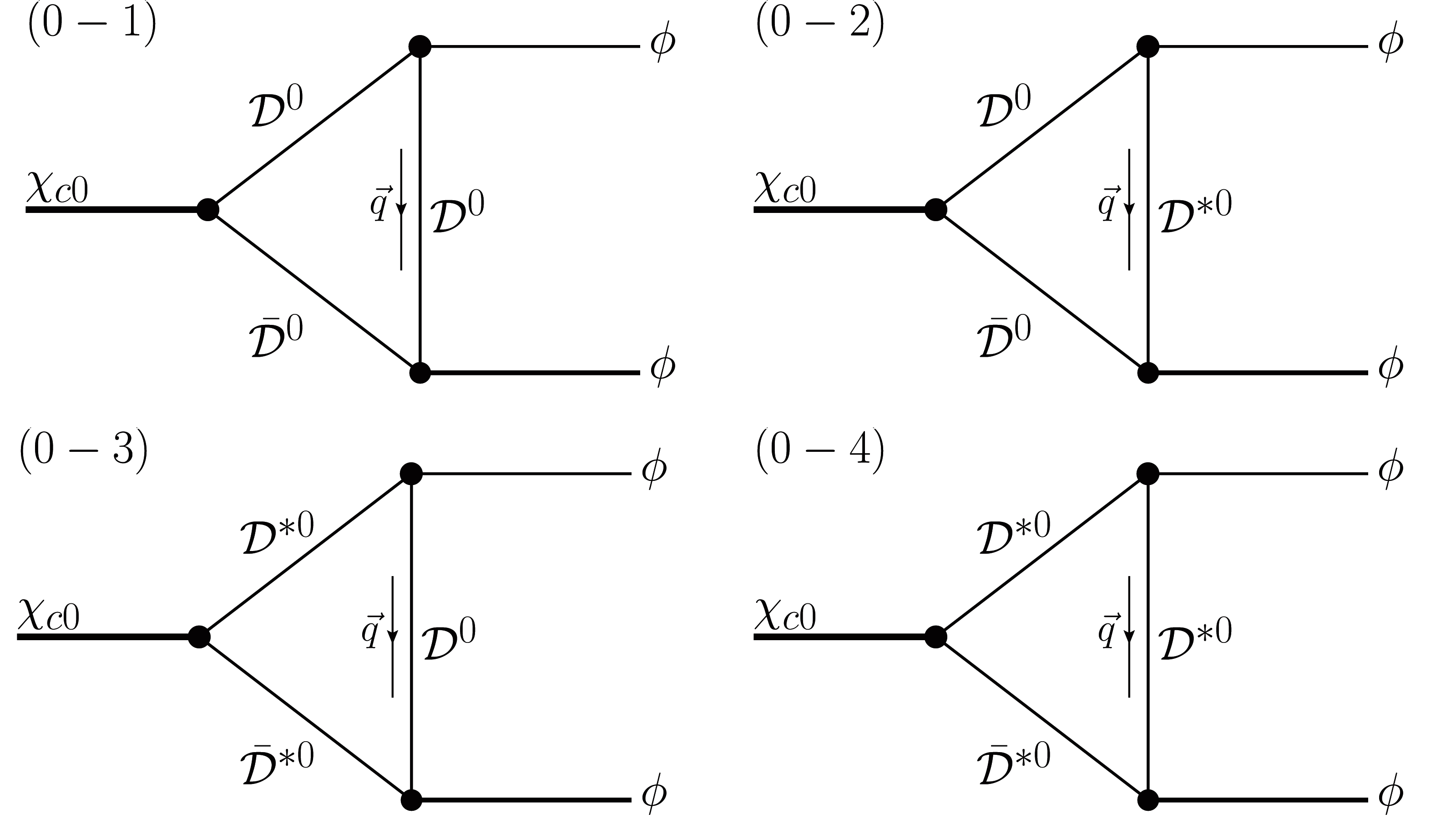}}
\caption{The Feynman diagrams depicting the $\chi_{c0}\to\phi\phi$ decay via $D$ meson loop.}
\label{fig:chic0-D-loop}
\end{figure}
\end{center}

\begin{center}
\begin{figure}[htbp]
\scalebox{0.075}{\includegraphics{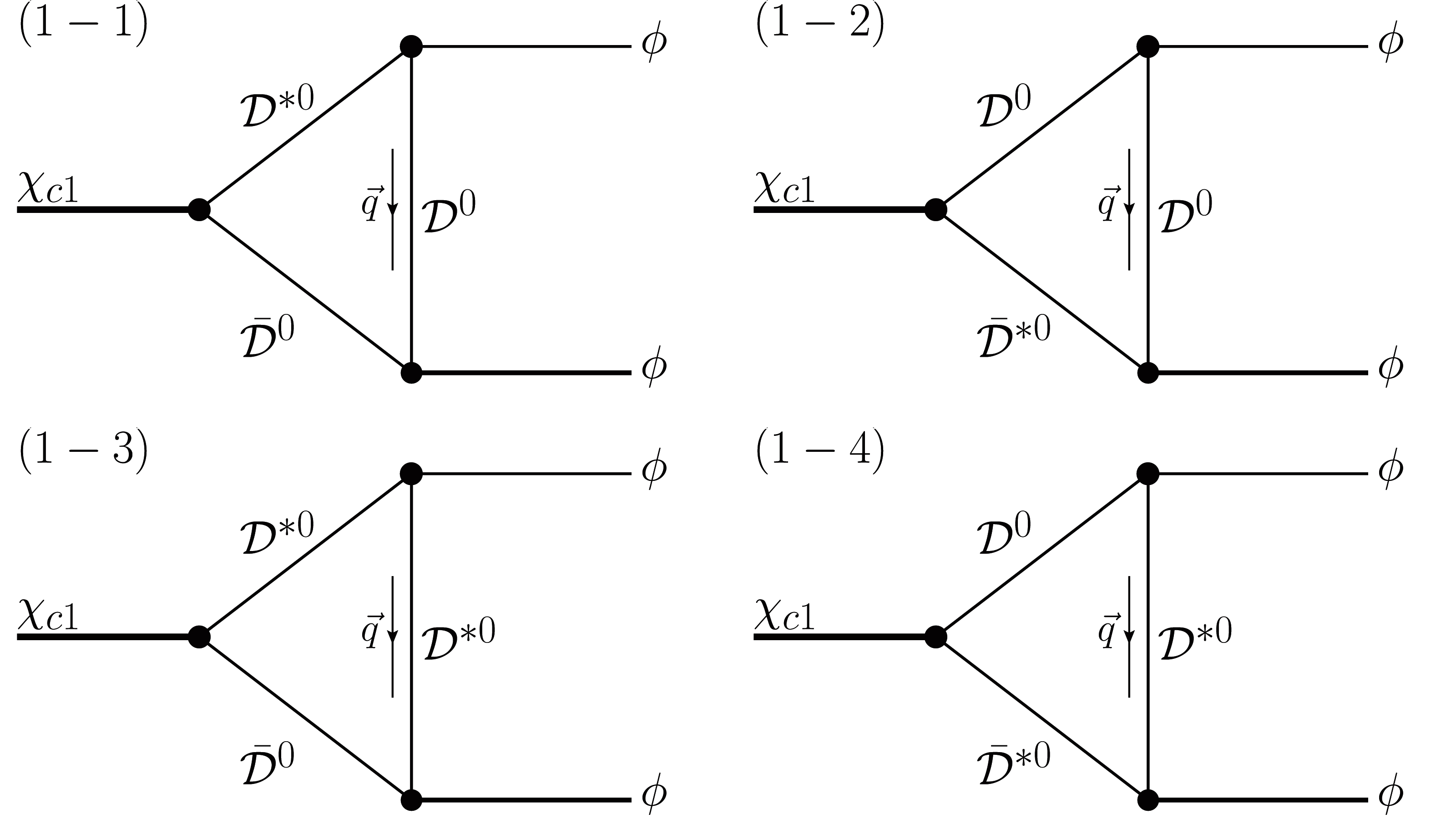}}
\caption{The Feynman diagrams depicting the $\chi_{c1}\to\phi\phi$ decay via $D$ meson loop.}
\label{fig:chic1-D-loop}
\end{figure}
\end{center}

\begin{center}
\begin{figure}[htbp]
\scalebox{0.075}{\includegraphics{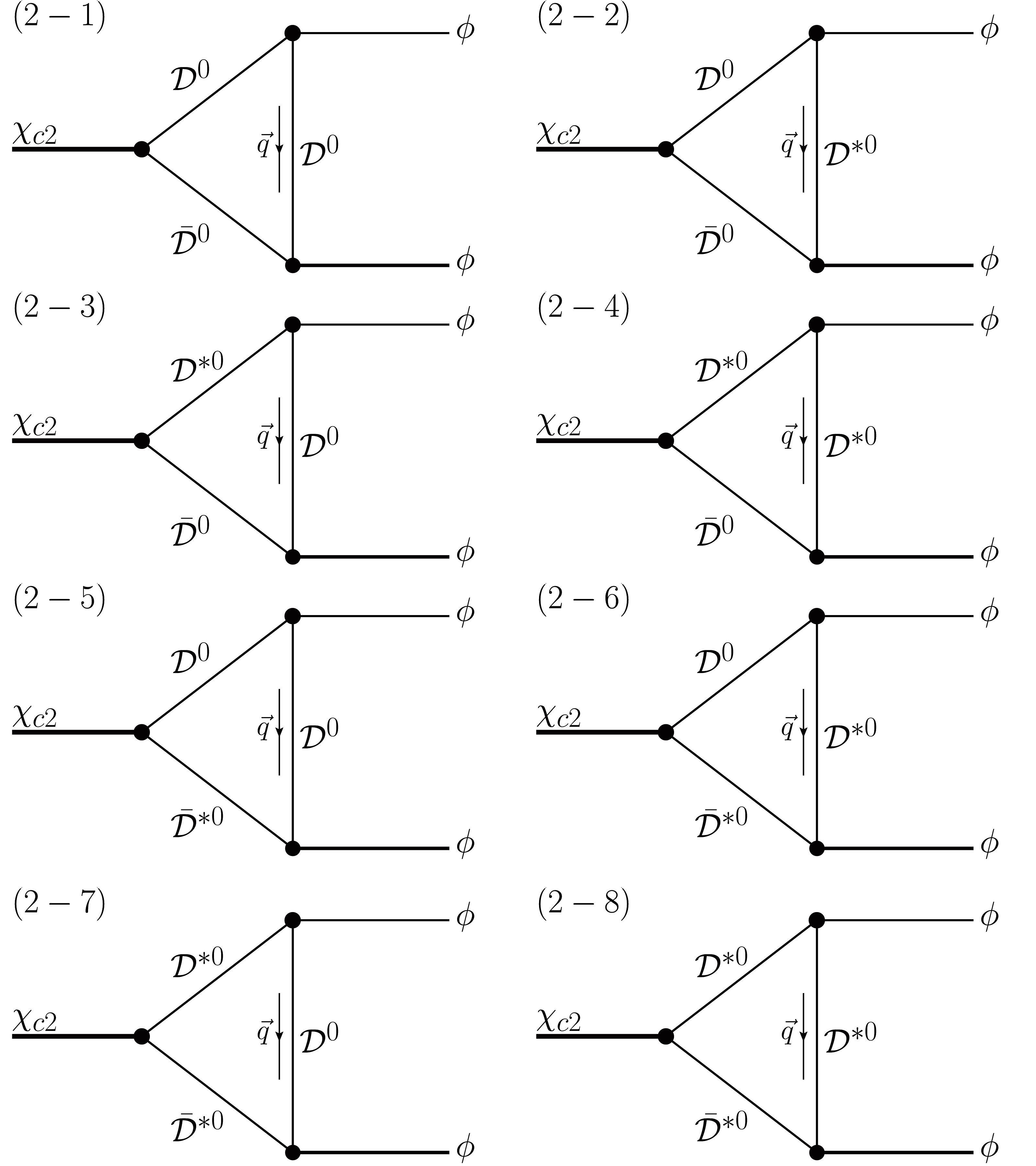}}
\caption{The Feynman diagrams depicting the $\chi_{c2}\to\phi\phi$ decay via $D$ meson loop.}
\label{fig:chic2-D-loop}
\end{figure}
\end{center}

Under the scenario of hadronic loop mechanism, the ${\chi_{cJ} \to \phi \phi}$ decays occur via the triangle loops composed of $D_{(s)}^{(\ast)}$ and $\bar{D}_{(s)}^{(\ast)}$, where these loops play the role of bridge to connect the initial $\chi_{cJ}$ and final states. In Figs. \ref{fig:chic0-D-loop}-\ref{fig:chic2-D-loop}, we present the Feynman diagrams depicting the ${\chi_{cJ} \to \phi \phi}$ decays

To calculate the decay amplitudes shown in Fig. \ref{fig:chic0-D-loop}-\ref{fig:chic2-D-loop}, we adopt the effective Lagrangian approach, thus at first we introduce the effective Lagrangians relevant to our calculation. For the interaction between $\chi_{cJ}$ and a pair of heavy-light mesons, the general form of the effective Lagrangian can be constructed under the chiral and heavy quark limits \cite{Casalbuoni:1996pg}
\begin{eqnarray}
\mathcal{L}_p &=& ig_1 \mathrm{Tr} \left[P^{(Q\bar{Q}) \mu} \bar{H}^{(\bar{Q}q)} \gamma_\mu \bar{H}^{(Q\bar{q})}\right] +H.c.,\label{eqs:PHH}
\end{eqnarray}
where ${P^{(Q\bar{Q})}}$ and ${H^{(Q\bar{q})}}$ denote the P-wave multiplet of charmonia and $(\mathcal{D}, \mathcal{D}^\ast)$ doublet, respectively. Their detailed expressions, as shown in Ref. \cite{Casalbuoni:1996pg,Kaymakcalan:1983qq,Oh:2000qr,Colangelo:2002mj}, can be written as
\begin{eqnarray}
{P^{(Q\bar{Q})}}^\mu &=& \frac{1+\slashed{v}}{2} \Big[\chi_{c2}^{\mu \alpha} \gamma_\alpha +\frac{1}{\sqrt{2}}  \varepsilon^{\mu \alpha \beta \gamma} v_\alpha \gamma_\beta \chi_{c1 \gamma} \nonumber\\
&&+\frac{1}{\sqrt{3}} \big(\gamma^\mu -v^\mu \big) \chi_{c0} +h_c^\mu \gamma_5 \Big] \frac{1-\slashed{v}}{2},\\
H^{(Q\bar{q})}&=&\frac{1+ \slashed{v}}{2} \left[\mathcal{D}^\ast_\mu \gamma^\mu-\mathcal{D} \gamma^5\right],
\end{eqnarray}
respectively, with definitions $\mathcal{D}^{(*)\dag} = (D^{(*)+},D^{(*)0},D_s^{(*)0})$ and $\mathcal{D}^{(*)} = (D^{(*)-},\bar{D}^{(*)0},\bar{D}_s^{(*)0})^T$. $H^{(\bar{Q}q)}$ corresponds to the doublet formed by homologous heavy-light anti-mesons, which can be obtained by applying the charge conjugation operation to $H^{(Q\bar{q})}$.

For the interaction between a light vector meson and two heavy-light mesons, the general form of the Lagrangian reads as \cite{Casalbuoni:1996pg,Cheng:1992xi,Yan:1992gz,Wise:1992hn,Burdman:1992gh,Falk}
\begin{eqnarray}\label{eqs:HHV}
\mathcal{L}_\mathcal{V} = i\beta \mathrm{Tr}[H^j v^\mu (-\rho_\mu)_j^i \bar{H}_i] + i\lambda \mathrm{Tr}[H^j \sigma^{\mu\nu} F_{\mu\nu} (\rho) \bar{H}_i],
\end{eqnarray}
where
\begin{eqnarray}
%\rho_\mu &=& i \frac{g_V}{\sqrt{2}} \hat{\rho}_\mu,\\
\rho_\mu &=& i \frac{g_V}{\sqrt{2}} \mathcal{V}_\mu,\\
F_{\mu\nu} (\rho) &=& \partial_\mu \rho_\nu - \partial_\nu \rho_\mu + [\rho_\mu,\rho_\nu],
\end{eqnarray}
%and ${\hat{\rho}_\mu}$ is a matrix of a vector octet $\mathcal{V}$, which has the form
and a vector octet $\mathcal{V}$ has the form \cite{Chen:2009ah}
\begin{small}
\begin{eqnarray}
\mathcal{V} &=&
 \left(
 \begin{array}{ccc}
\frac{\rho^0}{\sqrt{2}}+\kappa\omega^p+\zeta\phi^p & \rho^{+} & K^{\ast+}\\
\rho^{-} & -\frac{\rho^0}{\sqrt{2}}+\kappa\omega^p+\zeta\phi^p &  K^{\ast0}\\
 K^{\ast-} & \bar{K}^{\ast0} & \delta\omega^p+\sigma\phi^p
 \end{array}
 \right),\label{eqs:octet}
\end{eqnarray}
\end{small}
with
\begin{equation}
\label{eqs:mixing}
 \begin{split}
\kappa&=\frac{\cos \theta}{\sqrt{2}},\quad\quad\zeta=\frac{\sin \theta}{\sqrt{2}},\\
\delta&=\sin \theta,\quad\sigma=-\cos \theta.
\end{split}
\end{equation}

By expanding the Lagrangians in Eqs. (\ref{eqs:PHH}) and (\ref{eqs:HHV}), we get the following explicit forms of Lagrangians
\begin{eqnarray}\label{eqs:chicjDD}
&&\mathcal{L}_{\chi_{cJ} \mathcal{D}^{(\ast)} \mathcal{D}^{(\ast)}}\nonumber\\
&&= - g_{\chi_{c0} \mathcal{D} \mathcal{D} } \chi_{c0} \mathcal{D}
\mathcal{D}^\dagger - g_{\chi_{c0} \mathcal{D}^\ast
\mathcal{D}^\ast} \chi_{c0} \mathcal{D}_{\mu}^\ast \mathcal{D}^{\ast
\mu\dagger } \nonumber\\ && \quad +i g_{\chi_{c1} \mathcal{D}
\mathcal{D}^\ast} \chi_{c1}^\mu ( \mathcal{D}^{\ast }_\mu
\mathcal{D}^\dagger - \mathcal{D} \mathcal{D}^{\ast \dagger}_\mu )
\nonumber\\
&&\quad- g_{\chi_{c2} \mathcal{D} \mathcal{D}}
\chi_{c2}^{\mu \nu}
\partial_\mu \mathcal{D} \partial_\nu \mathcal{D}^\dagger\nonumber\\
&&\quad+ g_{\chi_{c2} \mathcal{D}^\ast \mathcal{D}^\ast} \chi_{c2}^{\mu
\nu}
(\mathcal{D}^\ast_{\mu} \mathcal{D}^{\ast \dagger}_\nu + \mathcal{D}^\ast_{\nu} \mathcal{D}^{\ast \dagger}_\mu) \nonumber\\
&&  \quad-ig_{\chi_{c2} \mathcal{D}^\ast \mathcal{D}} \varepsilon_{\mu
\nu \alpha \beta} \partial^\alpha \chi_{c2}^{\mu \rho}
(\partial_\rho \mathcal{D}^{\ast \nu} \partial^\beta
\mathcal{D}^\dagger -\partial^\beta \mathcal{D}
\partial_\rho \mathcal{D}^{\ast \nu \dagger} ),
\end{eqnarray}
\begin{eqnarray}\label{eqs:DDphi}
&&\mathcal{L}_{\mathcal{D}^{(\ast)}\mathcal{D}^{(\ast)} \mathcal{V}}
\nonumber\\&&= -ig_{\mathcal{D} \mathcal{D}\mathcal{V}} \mathcal{D}_i^\dagger
\lrpartial^\mu \mathcal{D}^j (\mathcal{V}_\mu)^i_j\nonumber\\
&&\quad-2 f_{\mathcal{D}^\ast \mathcal{D} \mathcal{V}} \varepsilon_{\mu \nu
\alpha \beta} (\partial^\mu
\mathcal{V}^\nu)^i_j (\mathcal{D}^\dagger_i \lrpartial^\alpha
\mathcal{D}^{\ast \beta j} -\mathcal{D}_i^{\ast \beta \dagger}
\lrpartial^\alpha \mathcal{D}^j) \nonumber\\&& \quad+ig_{\mathcal{D}^\ast
\mathcal{D}^\ast \mathcal{V}} \mathcal{D}^{\ast \nu \dagger}_i
\lrpartial^\mu \mathcal{D}^{\ast j}_\nu (\mathcal{V}_\mu)^i_j
\nonumber\\ &&\quad+4if_{\mathcal{D}^\ast \mathcal{D}^\ast \mathcal{V}}
\mathcal{D}^{\ast \dagger}_{i\mu} (\partial^\mu \mathcal{V}^\nu
-\partial^\nu \mathcal{V}^\mu)^i_j \mathcal{D}^{\ast j}_\nu.
\end{eqnarray}

With these Lagrangians given in Eq. (\ref{eqs:chicjDD}) and Eq. (\ref{eqs:DDphi}), the amplitudes of $\chi_{cJ}\to\phi\phi$ then can be written out. For $\chi_{c0} \to \phi \phi$ decay, with $\tilde{g}_{\mu\nu}(p) \equiv -g_{\mu\nu} + \frac{p_\mu p_\nu}{m_p^2}$, the amplitudes corresponding to Fig. \ref{fig:chic0-D-loop} are
\begin{eqnarray}\label{eqs:amp-begin}
\mathcal{M}_{(0-1)}&=&\int \frac{d^4 q}{(2\pi)^4} \frac{1}{k_1^2 - m_{\mathcal{D}}^2} \frac{1}{k_2^2 - m_{\mathcal{D}}^2} \frac{1}{q^2 - m_{\mathcal{D}}^2} \mathcal{F}^2(q^2)\nonumber\\
&&\times[-g_{\chi_{c0}\mathcal{D}\mathcal{D}}] [-g_{\mathcal{D} \mathcal{D} \phi} \epsilon^{\ast\zeta}_{\phi}(p_2)(k_{1\zeta}+q_\zeta)]\nonumber\\
&&\times [-g_{\mathcal{D} \mathcal{D} \phi} \epsilon^{\ast\lambda}_{\phi}(p_3)(q_\lambda-k_{2\lambda})],
\end{eqnarray}
\begin{eqnarray}
\mathcal{M}_{(0-2)}&=&\int \frac{d^4 q}{(2\pi)^4} \frac{1}{k_1^2 - m_{\mathcal{D}}^2} \frac{1}{k_2^2 - m_{\mathcal{D}}^2} \frac{\tilde{g}_{\xi\sigma}(q)}{q^2 - m_{\mathcal{D}^\ast}^2} \mathcal{F}^2(q^2)\nonumber\\
&&\times[-g_{\chi_{c0}\mathcal{D}\mathcal{D}}] [-2f_{\mathcal{D} \mathcal{D}^\ast \phi} \varepsilon^{\zeta\eta\kappa\xi} \epsilon^{\ast}_{\phi\zeta}(p_2) p_{2\eta} (k_{1\kappa}+q_\kappa)]\nonumber\\
&&\times[2f_{\mathcal{D} \mathcal{D}^\ast \phi} \varepsilon^{\lambda\rho\delta\sigma} \epsilon^{\ast}_{\phi\lambda}(p_3) p_{3\rho} (q_\delta-k_{2\delta})],
\end{eqnarray}
\begin{eqnarray}
\mathcal{M}_{(0-3)}&=&\int \frac{d^4 q}{(2\pi)^4} \frac{\tilde{g}_{\xi}^{\mu}(k_1)}{k_1^2 - m_{\mathcal{D}^\ast}^2} \frac{\tilde{g}_{\mu\sigma}(k_2)}{k_2^2 - m_{\mathcal{D}^\ast}^2} \frac{1}{q^2 - m_{\mathcal{D}}^2} \mathcal{F}^2(q^2)\nonumber\\
&&\times[-g_{\chi_{c0}\mathcal{D}^\ast \mathcal{D}^\ast}] [2f_{\mathcal{D} \mathcal{D}^\ast \phi} \varepsilon^{\zeta\eta\kappa\xi} \epsilon^{\ast}_{\phi\zeta}(p_2) p_{2\eta} (k_{1\kappa}+q_\kappa)]\nonumber\\
&&\times[-2f_{\mathcal{D} \mathcal{D}^\ast \phi} \varepsilon^{\lambda\rho\delta\sigma} \epsilon^{\ast}_{\phi\lambda}(p_3) p_{3\rho} (q_\delta-k_{2\delta})],
\end{eqnarray}
\begin{eqnarray}\label{eqs:amp-end}
\mathcal{M}_{(0-4)}&=&\int \frac{d^4 q}{(2\pi)^4} \frac{\tilde{g}^{\mu\psi}(k_1)}{k_1^2 - m_{\mathcal{D}^\ast}^2} \frac{\tilde{g}_{\mu}^{\iota}(k_2)}{k_2^2 - m_{\mathcal{D}^\ast}^2} \frac{\tilde{g}^{\gamma\upsilon}(q)}{q^2 - m_{\mathcal{D}^\ast}^2} \mathcal{F}^2(q^2)\nonumber\\
&&\times[-g_{\chi_{c0}\mathcal{D}^\ast \mathcal{D}^\ast}] [g_{\mathcal{D}^\ast \mathcal{D}^\ast \psi} g_{\eta \gamma}g_{\psi}^{\eta} (k_{1\zeta} + q_\zeta)\nonumber\\
&&- 4f_{\mathcal{D}^\ast \mathcal{D}^\ast \phi} p_2^\eta (g_{\gamma\eta}g_{\psi\zeta} - g_{\gamma\zeta}g_{\psi\eta})]\epsilon^{\ast\zeta}_{\phi}(p_2)\nonumber\\
&&\times [g_{\mathcal{D}^\ast \mathcal{D}^\ast \phi} g_{\rho \iota}g_{\upsilon}^{\rho} (q_\lambda-k_{2\lambda})- 4f_{\mathcal{D}^\ast \mathcal{D}^\ast \phi} p_3^\rho  \nonumber\\
&&\times (g_{\iota\rho}g_{\upsilon\lambda}- g_{\iota\lambda}g_{\upsilon\rho})]  \epsilon^{\ast\lambda}_{\phi}(p_3),
\end{eqnarray}

In the similar way, the amplitudes of $\chi_{c1} \to \phi \phi$ and $\chi_{c2} \to \phi \phi$ can be written out, which are collected into Appendix \ref{app:chic1} and Appendix \ref{app:chic2}, respectively.

In the calculation of the $\chi_{cJ} \to \phi \phi$ amplitudes, a dipole form factor, ${\mathcal{F}(q^2) = (m_E^2-\Lambda^2)^2 / (q^2-\Lambda^2)^2}$, is introduced to describe the structure effect and off-shell effect due to exchanging $\mathcal{D}^{(\ast)}$ mesons \cite{Chen:2009ah} at the $\mathcal{D}^{(\ast)}\mathcal{D}^{(\ast)}\phi$ and $\bar{\mathcal{D}}^{(\ast)}\mathcal{D}^{(\ast)}\phi$ vertices. This form factor plays a role similar to the Pauli-Villas renormalization scheme, which is often used to cancel the ultraviolet divergence in the loop integrals \cite{Itzykson:1980rh,Peskin:1995ev}. In the expression of $\mathcal{F}(q^2)$, $m_E$ is the mass of the exchanged $\mathcal{D}^{(\ast)}$ meson and $\Lambda$ denotes the cutoff, which is usually parameterized as ${\Lambda = m_E + \alpha_\Lambda \Lambda_{\mathrm{QCD}}}$. Here ${\Lambda_{\mathrm{QCD}} = 0.22~\mathrm{GeV}}$, $\alpha_\Lambda$ is a free parameter \cite{Liu:2006dq,Colangelo:2002mj,Chen:2011zv,Meng:2007tk,Meng:2008dd,Chen:2011qx,Chen:2014ccr,Chen:2011pv,Meng:2008bq,Liu:2009dr,Li:2013zcr,Cheng:2004ru}. Since the cutoff $\Lambda$ is required to be not too far away from the physical mass of the exchanged mesons, usually $\alpha_\Lambda$ should be around 1 \cite{Cheng:2004ru}. One notes that the multipole behavior of form factor was also suggested in the QCD sum rule study in Ref. \cite{Gortchakov:1995im}. In a series of published papers \cite{Chen:2011zv,Meng:2007tk,Meng:2008dd, Chen:2011qx,Chen:2014ccr,Chen:2011pv,Meng:2008bq,Liu:2006dq,Liu:2009dr,Li:2013zcr,Colangelo:2002mj,Cheng:2004ru}, this form factor was adopted in the study of transitions of charmonia, bottomonia, and $B$ decays. 

With Eqs. (\ref{eqs:amp-begin}-\ref{eqs:amp-end}), considering charge conjugation and isospin symmetries, the polarized amplitudes of ${\chi_{cJ} \to \phi \phi}$ read
\begin{eqnarray}
\mathcal{M}_J(i,\lambda_1,\lambda_2)=4 \sum\limits_{j} \mathcal{M}^q_{(J-j)}+2 \sum\limits_{j} \mathcal{M}^s_{(J-j)},
\end{eqnarray}
where $i$, $\lambda_1$ and $\lambda_2$ denote the helicities of $\chi_{cJ}$ and two $\phi$ mesons, respectively, $\mathcal{M}^q_{(J-j)}$ and $\mathcal{M}^s_{(J-j)}$ represent that the triangle loops are composed of charmed and charmed-strange mesons, respectively.

Thus, the helicity amplitudes can be calculated by the following expression
\begin{eqnarray}
|F_{\lambda_1,\lambda_2}^{(J)}|^2 = \sum\limits_{i}\rho_J(i)|\mathcal{M}_J(i,\lambda_1,\lambda_2)|^2,
\end{eqnarray}
where $\rho_J$ is the SDM given in Sec. \ref{sec:polar}, i.e.,
\begin{eqnarray}
\rho_0 &=& 1,\\
\rho_1 &=&\frac{1}{4} \mathrm{diag}\{1,2,1\}, \\
\rho_2 &=&\frac{3}{20} \mathrm{diag}\{2,1,\frac{2}{3},1,2\}.
\end{eqnarray}

Finally, the general expression of the decay widths of ${\chi_{cJ} \to \phi \phi}$ decays reads as
\begin{eqnarray}
\Gamma_{\chi_{cJ} \to \phi \phi} = \frac{1}{1+\delta} \frac{1}{ \sum\limits_{i} \rho_J(i) } \frac{1}{8\pi}  \frac{|\vec{p}_\phi|}{m^2_{\chi_{cJ}}} \sum\limits_{i,\lambda_1,\lambda_2} |\mathcal{M}_J(i,\lambda_1,\lambda_2)|^2,
\end{eqnarray}
where factor $\delta$ should be introduced if the final states are identical particles. Thus, for the discussed $\chi_{cJ}\to \phi \phi$ decays, we should take $\delta=1$.

\section{Numerical results}\label{sec:nr}

\subsection{Helicity amplitudes}\label{sec:nr-amp}

With the formula given in Sec. \ref{sec:loop}, we estimate all the helicity amplitudes $F^{(J)}_{\lambda_1,\lambda_2}$. Besides the masses taken from the Particle Data Group (PDG) \cite{Zyla:2020zbs}, other input parameters include the coupling constants, the mixing angle $\theta$ between $\omega^p$ and $\phi^p$, and the parameter $\alpha_\Lambda$ that appears in the expression of form factor $\mathcal{F}(q^2)$. For the coupling constants relevant to the interactions between $\chi_{cJ}$ and $D_{(s)}^{(*)}\bar{D}_{(s)}^{(*)}$, in the heavy quark limit, they are related to one gauge coupling constant $g_1$ given in Eq. (\ref{eqs:PHH}). By comparing Eq. (\ref{eqs:chicjDD}) and the expanded form of Eq. (\ref{eqs:PHH}) we can get
\begin{eqnarray}
  g_{\chi_{c0} \mathcal{DD}}&=&2\sqrt{3} g_1\sqrt{m_{\chi_{c0}}} m_{\mathcal{D}}, \ \
  g_{\chi_{c0} \mathcal{D}^\ast \mathcal{D}^\ast } =\frac{2}{\sqrt{3}}
  g_1 \sqrt{m_{\chi_{c0}}} m_{\mathcal{D}^\ast},\nonumber\\
  g_{\chi_{c1} \mathcal{D}\mathcal{D}^\ast} &=& 2\sqrt{2} g_1
  \sqrt{m_{\chi_{c1}} m_{\mathcal{D}}m_{\mathcal{D}^\ast}}, \ \
  g_{\chi_{c2} \mathcal{DD}} =2g_1 \frac{\sqrt{m_{\chi_{c0}}}
  }{m_{\mathcal{D}}}, \nonumber\\
  g_{\chi_{c2} \mathcal{D} \mathcal{D}^\ast} &=& g_1
  \sqrt{\frac{m_{\chi_{c2}}}{m_{\mathcal{D}^\ast}^3 m_{\mathcal{D}}}},
  \ \ g_{\chi_{c2} \mathcal{D}^\ast \mathcal{D}^\ast}
  =4g_1\sqrt{m_{\chi_{c2}}} m_{\mathcal{D}^\ast}.
\end{eqnarray}

We determine $g_1$ in the following way. For the interaction between the $\chi_{c0}$ state and a pair of $\mathcal{D}$ mesons, we define its matrix element as
\begin{eqnarray}
  \langle \mathcal{D}(p) \bar{\mathcal{D}}(p^\prime) | \chi_{c0} (P) \rangle = g_{\chi_{c0}\mathcal{D}\mathcal{D}}.
\end{eqnarray}
With the help of Isgur-Wise form factor \cite{Manohar:2000dt,Neubert:1993mb}, and the assumption that $\chi_{c0}$ gives the dominant contribution to the scalar current matrix element $\langle \mathcal{D}(p^\prime) | c\bar{c}|\mathcal{D}(p) \rangle$, together with the definition of the decay constant $f_{\chi_{c0}}$, i.e.,
\begin{eqnarray}
  \langle 0 | c\bar{c} | \chi_{c0}(P) \rangle = f_{\chi_{c0}} m_{\chi_{c0}},
\end{eqnarray}
in $(p-p^\prime)^2$-channel, we can express the scalar current matrix element $\langle \mathcal{D}(p^\prime) | c\bar{c}|\mathcal{D}(p) \rangle$ as
\begin{eqnarray}
  \langle \mathcal{D}(p^\prime) | c\bar{c}|\mathcal{D}(p) \rangle
  &=& \frac{g_{\chi_{c0}\mathcal{D}\mathcal{D}} f_{\chi_{c0}} m_{\chi_{c0}}}{(p-p^\prime)^2-m_{\chi_{c0}}^2} \nonumber\\
  &=& (1+w) \xi(w) m_\mathcal{D},
\end{eqnarray}
where $\xi(w)$ is the normalized Isgur-Wise form factor, i.e., $\xi(w=1)=1$ \cite{Manohar:2000dt,Neubert:1993mb}, $w=\frac{p \cdot p^\prime}{m_{\mathcal{D}}^2}$. Thus, when $p=p^\prime$, we have $w=1$, then we can relate $g_{\chi_{c0}\mathcal{D}\bar{\mathcal{D}}}$ to $f_{\chi_{c0}}$ as
\begin{eqnarray}
  g_{\mathcal{D}\mathcal{D}\chi_{c0}} = -2\frac{m_{\mathcal{D}}m_{\chi_{c0}}}{f_{\chi_{c0}}}.
\end{eqnarray}
So it is easy to see that $g_1=-\sqrt{m_{\chi_{c0}}\over3}\frac{1}{f_{\chi_{c0}}}$, with $f_{\chi_{c0}}=0.51$ GeV, which comes from an analysis of QCD sum rule \cite{Colangelo:2002mj,Chen:2010re}.

For the coupling constants of $D_{(s)}^{(*)}\bar{D}_{(s)}^{(*)}\phi$ interactions, they are determined from comparison of Eq. (\ref{eqs:DDphi}) to the expansion of Eq. (\ref{eqs:HHV}), i.e.
\begin{eqnarray}
  g_{D_s D_s \phi} &=&g_{D_s ^\ast D_s ^\ast \phi}=\frac{\beta g_V}{\sqrt{2}}\sigma,\nonumber\\
  f_{D_s D_s^\ast \phi} &=&\frac{f_{D_s^\ast D_s^\ast \phi}}{m_{D_s^\ast}} =\frac{\lambda g_V}{\sqrt{2}}\sigma,\nonumber\\
  g_{D D \phi} &=&g_{D ^\ast D ^\ast \phi}=\frac{\beta g_V}{\sqrt{2}}\zeta,\nonumber\\
  f_{D D^\ast \phi} &=&\frac{f_{D^\ast D^\ast \phi}}{m_{D^\ast}} =\frac{\lambda g_V}{\sqrt{2}}\zeta\nonumber
\end{eqnarray}
with $\beta=0.9$ and $\lambda=0.56 \ \mathrm{GeV}^{-1}$. The values of $\beta$ and $\lambda$ come from the vector meson dominance model and the analyses of the $B \to K^\ast \pi$ decays, respectively \cite{Isola:2003fh}. Additionally, we have $g_V=m_\rho/f_\pi$ associated with the pion decay constant $f_\pi=132$ MeV \cite{Cheng:1992xi, Yan:1992gz, Wise:1992hn,Burdman:1992gh}.

There still exists an undetermined parameter, i.e., the mixing angle $\theta$ between $\omega^p$ and $\phi^p$. Using the Gell-Mann-Okubo mass formula \cite{Colangelo:2002mj,Burakovsky:1997sd,Kucukarslan:2006wk}, this parameter is roughly determined to be $\theta=3.7\degree$. On the other side, in the chiral perturbation theory, this mixing angle is related to the vector meson decay widths, $\Gamma(\omega \to \pi^+ \pi^-)$, $\Gamma(\rho^0 \to \pi^+ \pi^-)$ and $\Gamma(\phi \to \pi^+ \pi^-)$. Using experimental results \cite{Zyla:2020zbs}, one got $\theta=(3.4\pm0.2)\degree$ \cite{Kucukarslan:2006wk}, which is consistent with $\theta=3.7\degree$. Thus, we set $\theta=(3.4\pm0.2)\degree$ \cite{Chen:2009ah,Kucukarslan:2006wk,Benayoun:1999fv,Dolinsky:1991vq} in the calculation of the helicity amplitudes $F^{(J)}_{\lambda_1,\lambda_2}$.

The parameter $\alpha_\Lambda$ can be determined using the experimental data for the branching ratios of $\chi_{cJ} \to \phi \phi$ decays \cite{Zyla:2020zbs}. In our calculation, we find that if the branching ratios $\mathcal{B}(\chi_{cJ} \to \phi \phi)$ given by PDG \cite{Zyla:2020zbs} are reproduced simultaneously, $\alpha_\Lambda$ should be taken within the interval [1.15,1.35]. The values obey the cutoff requirement, and leads to the $\Lambda$ close to the physical mass of the exchanged mesons \cite{Cheng:2004ru}. In addition, they are consistent with those in Ref. \cite{Chen:2009ah}.

With the above parameter settings, we find that the ratios between helicity amplitudes are insensitive to the $\alpha_\Lambda$ and $\theta$, and the ratios change a little with variations of $\alpha_\Lambda \in [1.15,1.35]$ and $\theta=(3.4\pm0.2)\degree$. We get
\begin{eqnarray}
  \left|\frac{F^{(0)}_{1,1}}{F^{(0)}_{0,0}}\right| =0.359 \pm 0.019,
\end{eqnarray}
for the $\chi_{c0} \to \phi \phi$ decay,
\begin{eqnarray}
  \left|\frac{F^{(1)}_{1,0}}{F^{(1)}_{0,1}}\right|= 1,&
  \left|\frac{F^{(1)}_{1,1}}{F^{(1)}_{0,1}}\right|= 0,
\end{eqnarray}
for the $\chi_{c1} \to \phi \phi$ decay, and
\begin{eqnarray}
  \left|\frac{F^{(2)}_{1,0}}{F^{(2)}_{0,0}}\right| = \left|\frac{F^{(2)}_{0,1}}{F^{(2)}_{0,0}}\right| &=& 1.285 \pm 0.017,\\
  \left|\frac{F^{(2)}_{1,-1}}{F^{(2)}_{0,0}}\right| = \left|\frac{F^{(2)}_{-1,1}}{F^{(2)}_{0,0}}\right| &=& 5.110 \pm 0.057,\\
  \left|\frac{F^{(2)}_{-1,-1}}{F^{(2)}_{0,0}}\right| = \left|\frac{F^{(2)}_{1,1}}{F^{(2)}_{0,0}}\right| &=& 0.465 \pm 0.002.
\end{eqnarray}
for the $\chi_{c2} \to \phi \phi$ case.

The ratios for the $\chio$ decays are equivalent to those determined by the parity conservation and identical particle requirement. Especially, it is interesting to note that the ratios of $F^{(2)}_{1,0}$ and $F^{(2)}_{1,-1}$ are larger than one. This indicates they receive some contributions from the long distance effects, modeled by the rescattering of charm mesons in the $\chit\to\ff$ decays. We expect the measurement available in the future, and used for testing the hadron loop mechanism.

\subsection{Polarization observables}

Apart from the directly measurements on the ratios given in Sec. \ref{sec:nr-amp}, the $t_{ij}$ moments, $\langle t_{ij}\rangle$, can also be selected as the spin observables, since their distributions are directly related to the helicity amplitude $F^{(J)}_{\lambda_1,\lambda_2}$. The $t_{ij}$ observables are constructed only with the Kaon angles in $\phi$ decays. Thus, the $\langle t_{ij} \rangle$ moments should be independent on any parameter from theoretical investigations. In experiment, the $\langle t_{ij} \rangle$ moments are defined as
\begin{equation}
\langle t_{ij} \rangle = {1\over I_0}\int t_{ij} |\mathcal{M}|^2 d\Omega_2d\Omega_3,
\end{equation}
where $|\mathcal{M}|^2$ denotes the joint angular distribution for the $\chicj\to\ff\to2(\kk)$ decay and $d\Omega_i=d\cos\theta_id\phi_i(i=2,3)$ is the angles to be integrated out. $I_0$ is the normalization factor.

One exception is the $\chiz\to\ff$ decay, in which the multipole parameters $C_{ij}$ are independent on the angles of $\theta_1$ or $\phi_1$. Thus, the $\langle t_{ij} \rangle$ moments are uniformly distributed, and they can not be used as observable. Instead, we chose an observable $\mu=\sin^2\theta_2\sin^2\theta_3$ to express two $\phi$ spin entanglements produced from the $\chiz$ decays. With the joint angular distribution $\mathcal{W}_0$, one has
\begin{equation}
\langle \mu\rangle \propto 1 + 16|F^{(0)}_{1,1}/F^{(0)}_{0,0}|^2 \cos^2(\phi_2+\phi_3).
\end{equation}

\begin{figure}[htbp]
\centering
\mbox{\begin{overpic}[scale=0.4]{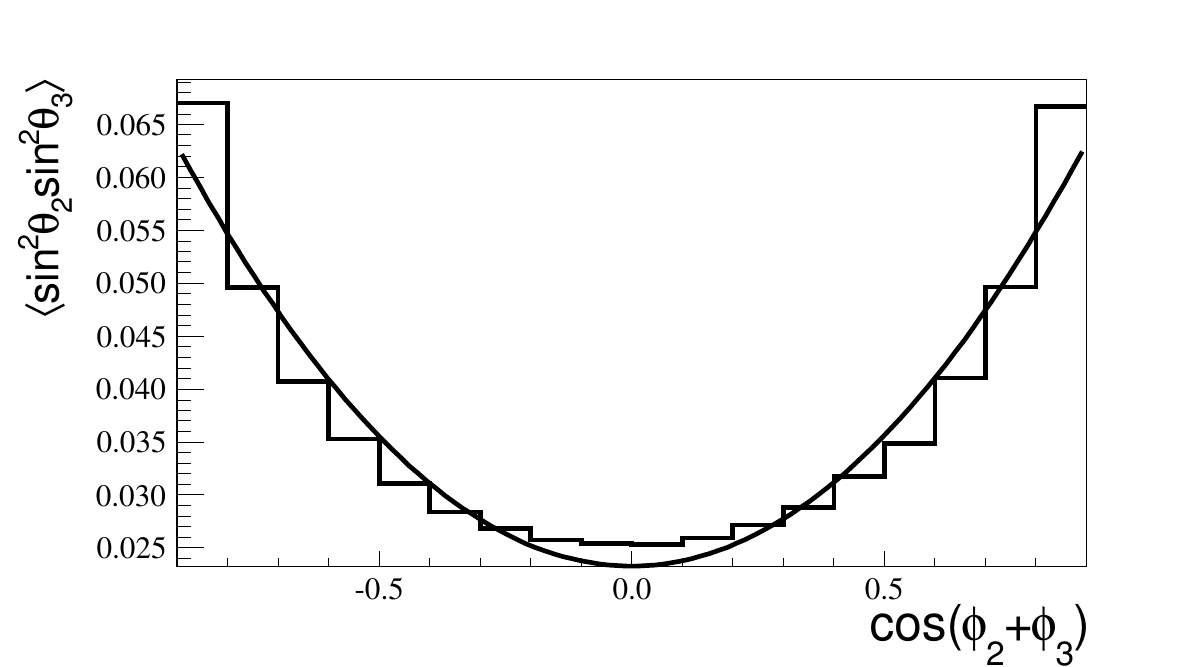} \end{overpic} }
{\caption{Distribution of $\langle \sin^2\theta_2\sin^2\theta_3 \rangle $ moment versus $\cos(\phi_2+\phi_3)$  for the $\chiz\to\ff\to 2(\kk)$. Histogram is filled with the MC events, and the curve shows the distribution of $1+2 \cos^2(\phi_2+\phi_3)$.}\label{chic0t00}}
\end{figure}

An ensemble of events is generated by using the $\chiz$ decay amplitude $\mathcal{W}_0$. And the ratio of amplitude is fixed to the central value of calculation, namely, $|F^{(0)}_{1,1}/F^{(0)}_{0,0}|=0.359$. The $\langle \sin^2\theta_2\sin^2\theta_3 \rangle $ moment of these TOY Monte-Carlo (MC) events is shown in Fig. \ref{chic0t00}. One can see that the MC distribution is consistent with the expectation of $1+2 \cos^2(\phi_2+\phi_3)$.

For the $\chio\to\ff$ decay, it conserves parity and the decay amplitude respects the identical particle symmetry when exchanging two $\phi$ mesons. Thus, the helicity amplitudes are able to factor out as an overall factor in the angular distribution. The $\phi$ angular distribution is independent on the amplitudes, and it is reduced to
\begin{equation}\label{eq:chic1:theta}
{dN\over d\cos\theta_1}\propto 1-{1\over 3} \cos^2\theta_1,
\end{equation}
which corresponds to the observation of moment $\langle t_{00}\rangle$ for the $\chio\to\ff$ decay.

\begin{figure}[htbp]
\centering
\mbox{\begin{overpic}[scale=0.4]{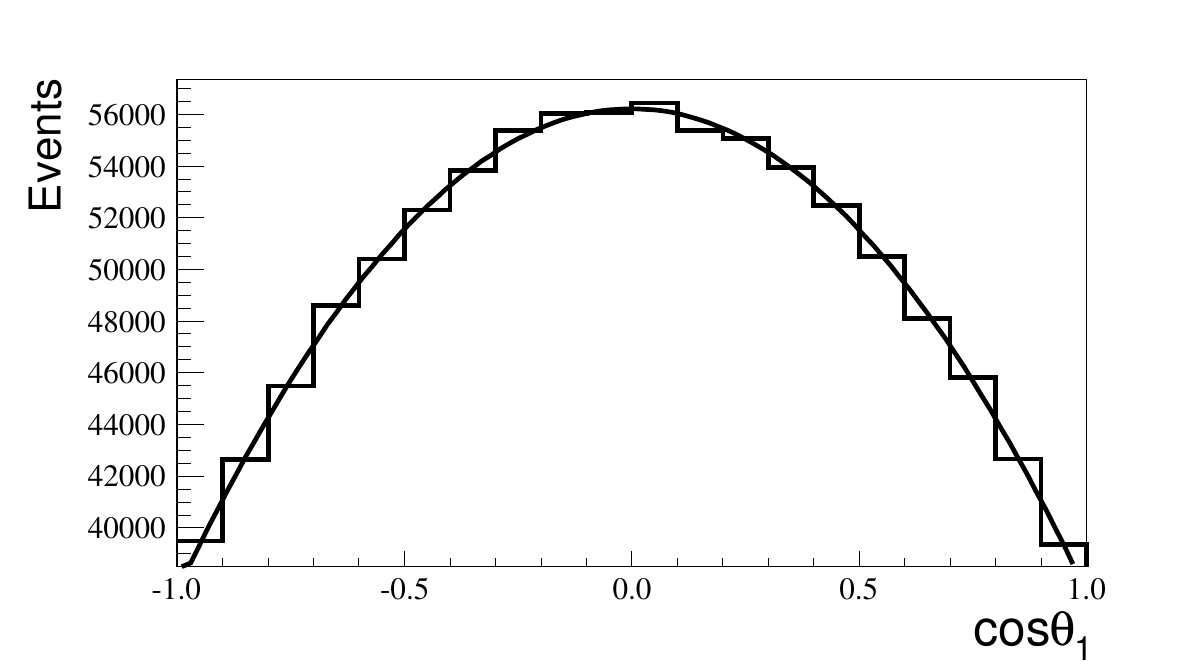} \end{overpic} }
{\caption{Angular distribution of $\phi$ meson in $\chio$ decays. Histogram is filled with the MC events, and the curve shows the distribution of $1-{1\over3} \cos^2\theta_1$.}\label{chic1costheta1}}
\end{figure}

We generate an ensemble of MC events for the $\chio$ decay with the amplitudes constrained by the requirements of parity conservation and the identical particle symmetry, namely, $F^{(1)}_{1,1}=0, ~F^{(1)}_{1,0}=-F^{(1)}_{0,1}$. Figure \ref{chic1costheta1} shows the angular distribution for the $\phi$ meson from the $\chio$ decays. One can see that the distribution is well consistent with the expected one as given by Eq. (\ref{eq:chic1:theta}).

One significant feature of $t_{ij}$ moments for $\chio$ decays is that their distributions are well determined only with the fundamental conservation rule and symmetry relations, being independent on the helicity amplitudes $F^{(1)}_{\lambda_1,\lambda_2}$.  For example, some $\langle t_{ij}\rangle$ moments are determined to be
\begin{eqnarray}
\langle t_{55}\rangle &\propto& 1-{1\over 2} \cos^2\theta_1,\\
\langle t_{60}\rangle ,\langle t_{06}\rangle,\langle t_{66}\rangle&\propto& 1-{1\over 3} \cos^2\theta_1,\\
\langle t_{80}\rangle,\langle t_{08}\rangle,\langle t_{68}\rangle,\langle t_{86}\rangle &\propto&1-~\cos^2\theta_1.
\end{eqnarray}

\begin{figure}[htbp]
\centering
\mbox{\begin{overpic}[scale=0.4]{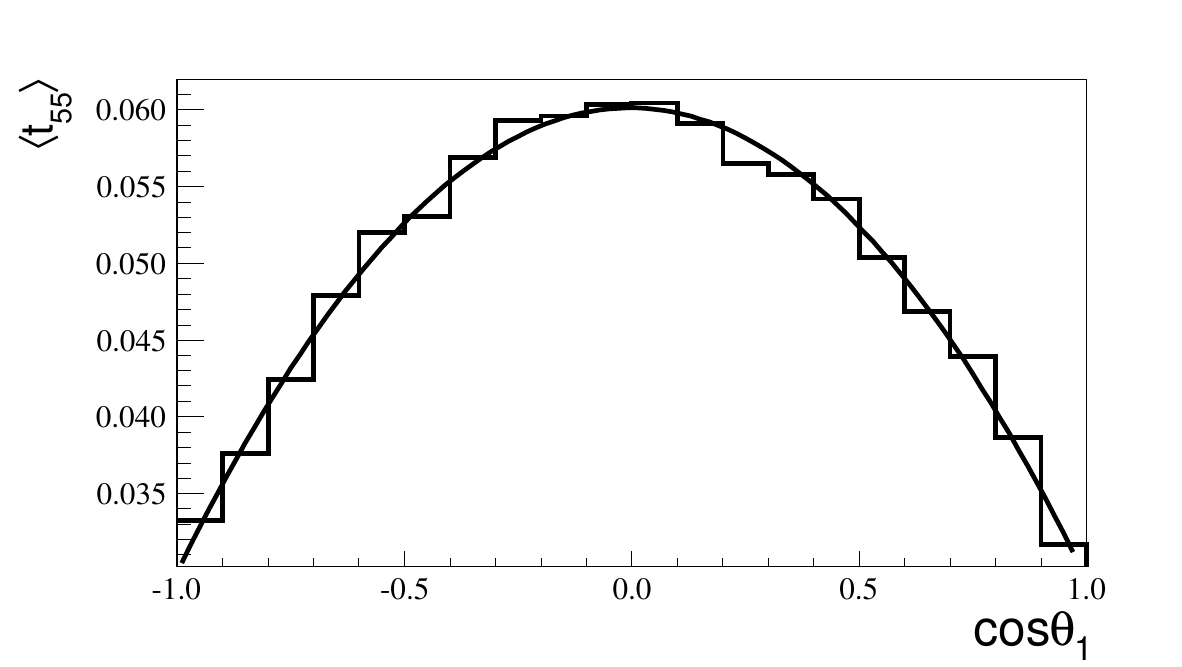} \end{overpic} }
{\caption{Moment distribution of $\langle t_{55}\rangle$ for $\chio$ decays. Histogram is filled with the MC events, and the curve shows the distribution of $1-{1\over 2} \cos^2\theta_1$.}\label{chic1t55}}
\end{figure}

Figure \ref{chic1t55} shows the $\langle t_{55}\rangle$ moment distribution filled with the $\chio$ MC events. The curve shows the expected distribution, and it is well consistent with the MC events.

\begin{figure}[htbp]
\centering
\mbox{\begin{overpic}[scale=0.4]{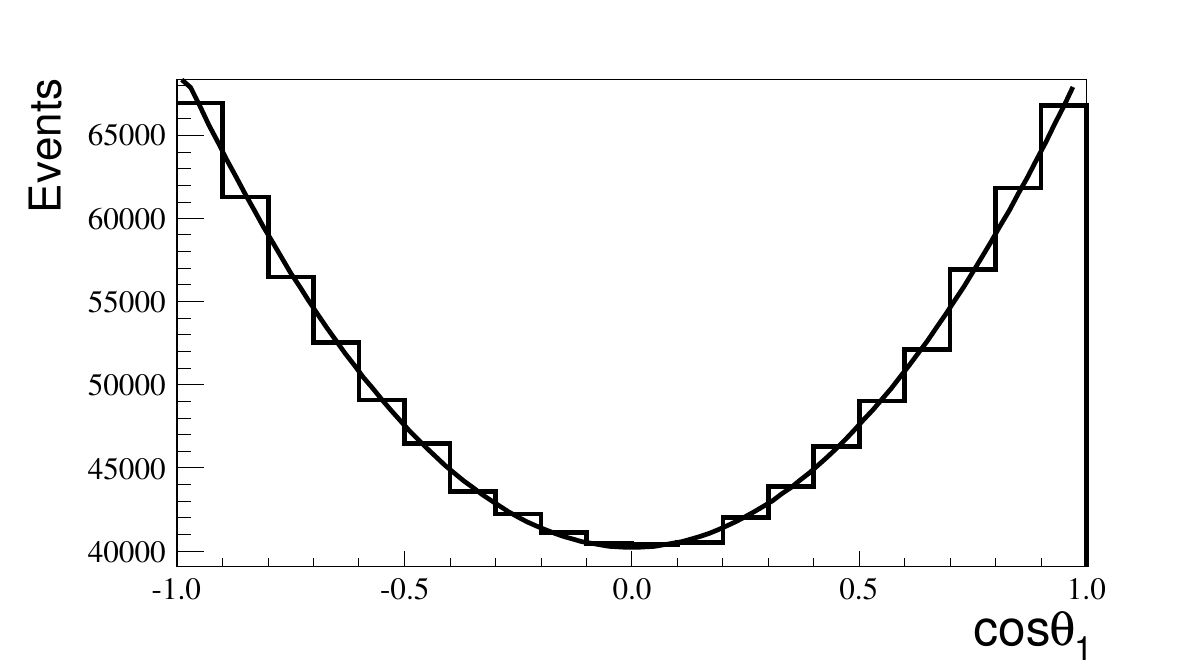} \end{overpic} }
{\caption{Angular distribution of the $\phi$ meson for $\chit$ decays. Histogram is filled with the MC events, and the curve shows the distribution of $1+0.736 \cos^2\theta_1$.}\label{chic2costheta}}
\end{figure}

To show the $\langle t_{ij}\rangle$ moments for the $\chit\to\ff\to2(\kk)$ decay, we generated MC events with the central values of predicted amplitude ratios, i.e. $|F^{(2)}_{1,0}|/|F^{(2)}_{0,0}|=|F^{(2)}_{0,1}|/|F^{(2)}_{0,0}|=1.285,~|F^{(2)}_{1,-1}|/|F^{(2)}_{0,0}|=|F^{(2)}_{-1,1}|/|F^{(2)}_{0,0}|=5.11$ and $|F^{(2)}_{1,1}|/|F^{(2)}_{0,0}|=0.465$. The $\langle t_{00}\rangle$ moments corresponds to the $\phi$ meson angular distribution. It reads as
\begin{equation}
{dN\over d\cos\theta_1}\propto 1 + \alpha \cos^2\theta_1
\end{equation}
with the angular distribution parameter
\begin{eqnarray}\alpha=-\frac{3 \left[|F^{(2)}_{0,0}|^2+2 \left(-|F^{(2)}_{1,-1}|^2+|F^{(2)}_{1,0}|^2+|F^{(2)}_{1,1}|^2\right)\right]}{5 |F^{(2)}_{0,0}|^2+6
|F^{(2)}_{1,-1}|^2+18 |F^{(2)}_{1,0}|^2+10 |F^{(2)}_{1,1}|^2}.
\end{eqnarray}
Using the ratios, one has $\alpha=0.736$.
Figure \ref{chic2costheta} shows the angular distribution (histogram) for the $\phi$ meson filled with the MC events, and the comparison with the predicted angular distribution (curve).

\begin{figure}[htbp]
\centering
\mbox{\begin{overpic}[scale=0.4]{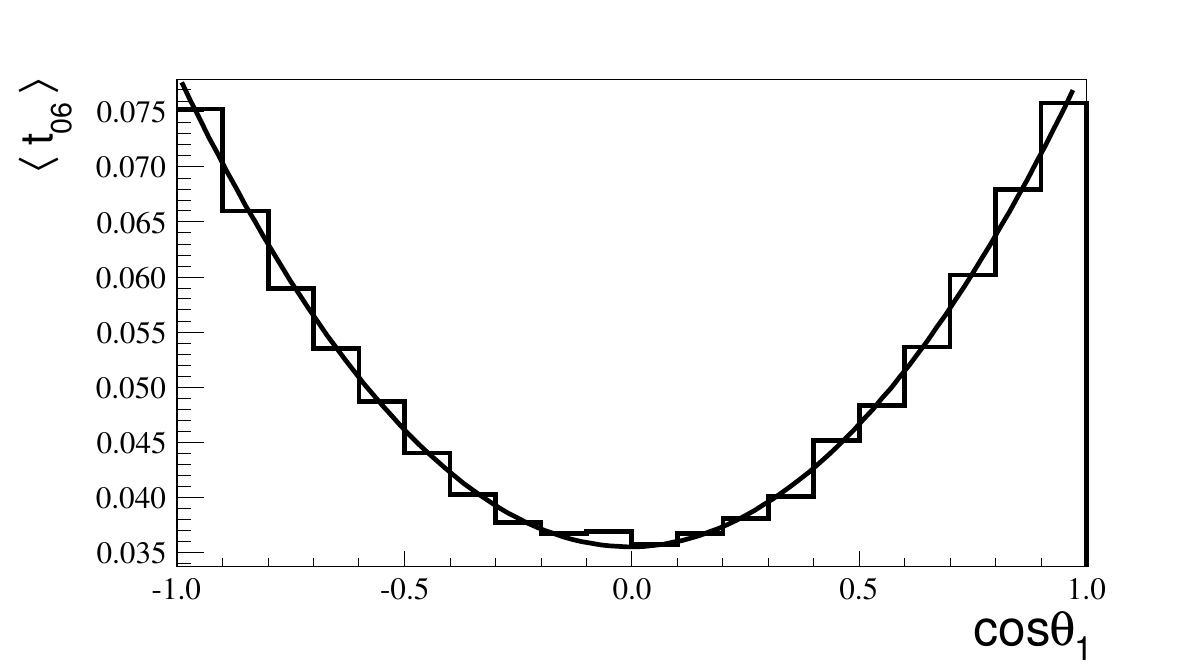} \end{overpic} }
{\caption{Angular distribution of the $\phi$ meson for $\chit$ decays. Histogram is filled with the MC events, and the curve shows the distribution of $1+1.26 \cos^2\theta_1$.}\label{chic2t06}}
\end{figure}

Another moment, $\langle t_{06}\rangle$ or $\langle t_{60}\rangle$, can also be used to reveal the ratios. It distributes with the form
$\langle t_{06}\rangle\propto 1+\alpha_1\cos^2\theta_1$ with
\begin{equation}
\alpha_1=-\frac{3 \left(2 |F^{(2)}_{0,0}|^2+2 |F^{(2)}_{1,-1}|^2+|F^{(2)}_{1,0}|^2-2 |F^{(2)}_{1,1}|^2\right)}{10 |F^{(2)}_{0,0}|^2-6|F^{(2)}_{1,-1}|^2+9 |F^{(2)}_{1,0}|^2-10 |F^{(2)}_{1,1}|^2}.
\end{equation}
Using the predicted ratios, we get $\alpha_1=1.24$. Figure \ref{chic2t06} shows the $\langle t_{06}\rangle$ distribution, filled with the MC events, which is comparable with the predicted distribution with $\alpha_1=1.24$.

\begin{figure}[htbp]
\centering
\mbox{\begin{overpic}[scale=0.4]{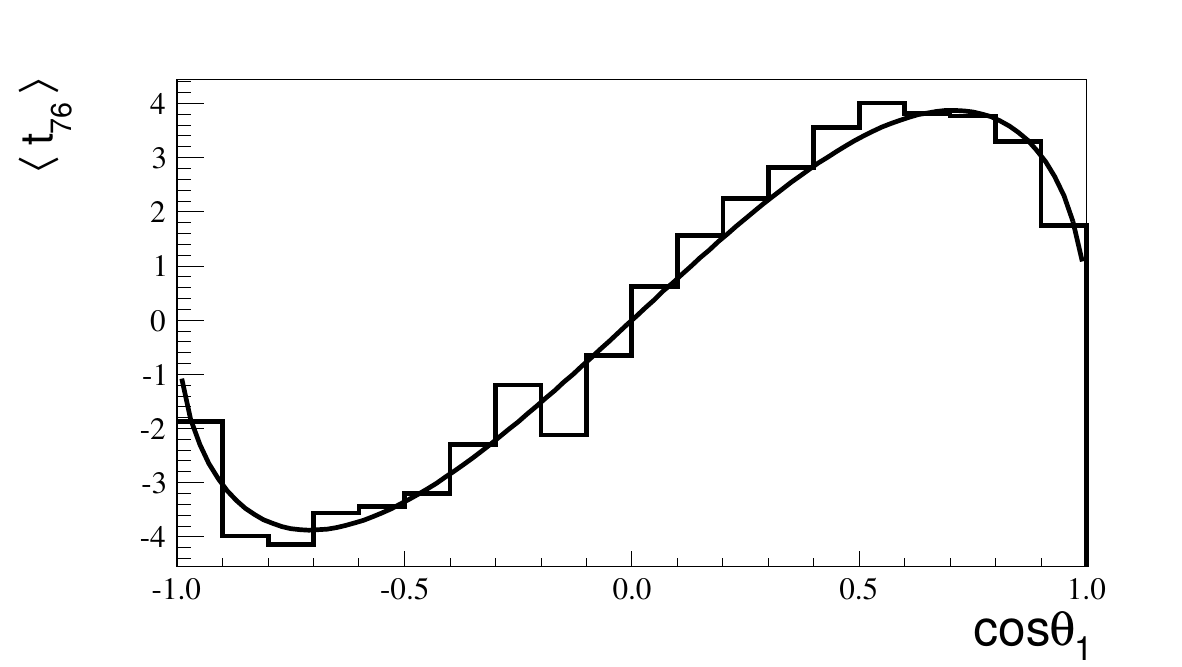} \end{overpic} }
{\caption{Distribution of the $\langle t_{76}\rangle$ moment for $\chit$ decays. Histogram is filled with the MC events, and the curve shows the distribution of $x\sqrt{1-x^2}$ with $x=\cos\theta_1$.}\label{chic2t76}}
\end{figure}

There are some $\langle t_{ij}\rangle$ moments in the $\chit$ decays, distributing independently on the amplitude ratios. After factoring out the amplitudes, we obtain these moment distributions versus $x=\cos\theta_1$, i.e.,
\begin{eqnarray}
\langle t_{44}\rangle &\propto& 1-6x^2/10,\\
\langle t_{76}\rangle,\langle t_{67}\rangle &\propto& x\sqrt{1-x^2},\\
\langle t_{80}\rangle,\langle t_{08}\rangle &\propto& 1-x^2.
\end{eqnarray}
Figure \ref{chic2t76} shows the $\langle t_{76}\rangle$ distribution, for example, for $\chit$ decays, and the comparison with the predicted one.

\section{Summary}\label{sec:summary}

The anomalous decay widths of the $\chi_{cJ} \to VV~(VV=\omega\omega,~\omega\phi~\mathrm{and}~\phi\phi)$ decays \cite{Ablikim:2018ogu,Ablikim:2011aa} indicate that the non-perturbative effect of strong interaction cannot be ignored. For accommodating the nonperturbative effect of strong interaction, the HLM scenario is adopted to study the branching ratios of $\chi_{cJ} \to VV~(VV=\omega\omega,~\omega\phi~\mathrm{and}~\phi\phi)$ decay \cite{Liu:2009vv,Chen:2009ah}. Although the measured branching ratios of the $\chi_{c1} \to \omega \omega$ and $\chi_{c1} \to \phi \phi$ decays can be reproduced well, it is not the end of whole story. In this work, we motivate to look for additional evidence to reveal the hadronic loop mechanism underlying 
the $\chi_{cJ} \to \phi \phi$ decays \cite{Ablikim:2018ogu,Ablikim:2011aa}.

Inspired by Refs. \cite{Chen:2009ah,Chen:2012ih,Chen:2013gka,BESIII:2016dda}, we propose that the analysis of polarization transfer in the $\chi_{cJ} \to VV$ decay can be applied to probe the hadronic loop mechanism, which becomes a main task of this work. One advantage of choosing the $\chicj\to\phi\phi$ decay is due to the fact that the two $\phi$ decays can provide rich spin observables. Another advantage is that these decays are accessible in experiment with high detection efficiency and two $\phi$ mesons can be cleanly reconstructed with low level backgrounds. A high statistics allows one to perform the angular distribution analyses and get the information of the $\phi$ polarization, which can shed light on the underlying decay mechanism of the $\chi_{cJ} \to VV$ decays.

In the scenario of hadronic loop mechanism, we find that the ratios of the helicity amplitudes for the $\chi_{cJ} \to \phi \phi$ decays is less dependent on the $\theta$ and $\alpha_\Lambda$, which are the mixing angle and a free parameter in the form factor respectively. Thus, we think that these ratios are the characteristic of HLM, and can be chosen as the observables. If they are measured in the future, the hadronic loop mechanism can be tested unambiguously.

In addition, by using the predicted amplitude ratios, we show that the observation of moments $\langle t_{ij} \rangle$ can be used to manifest the nontrivial polarization. For the $\chiz$ decays, the choice of the spin observable is quite limited due to the fact that the total spin of the $\ff$ system is constrained to be zero. Thus, the spins of two $\phi$ mesons are antiparallel for the $\chiz$ decays. For the $\chio\to\ff$ decays, the helicity amplitudes can be well determined by considering parity conservation and the identical particle symmetry. For the $\chit\to\ff$ decays, the abundant information of the $\ff$ spin configurations allows us to directly detect the helicity amplitudes from the observation of different $\langle t_{ij}\rangle$ moments. The patterns of these moments are presented based on the predicted amplitude ratios, which can be tested by expeirment in the near future.

In 2019, BESIII released the white paper on its future physics program \cite{Ablikim:2019hff}. With the accumulation of charmonium data, we suggest that BESIII should pay more attentions to the study of polarization of the corresponding decays, which may provide extra information for us to reveal the underlying decay mechanism of charmonia. Obviously, the present work provides a
typical example and a new task for experiment.

\section*{Acknowledgments}
This work is supported by the National Natural Science Foundation of China under Grant No. 11875262, 12047501 and 11835012, and the China National Funds for Distinguished Young Scientists under Grant No. 11825503, National Key Research and Development Program of China under Contract No. 2020YFA0406400, the 111 Project under Grant No. B20063.

\newpage
\begin{appendices}
\section{Spin observable $t_{ij}$.}\label{appA}
The obtained spin observables $t_{ij}$ are
\begin{eqnarray}
t_{00} &=&{1\over 9},\nonumber\\
t_{40}&=& -\frac{\sin ^2\left(\theta _2\right) \sin \left(2 \phi _2\right)}{3 \sqrt{3}},\nonumber\\
t_{04}&=& -\frac{\sin ^2\left(\theta _3\right) \sin \left(2 \phi _3\right)}{3 \sqrt{3}},\nonumber\\
t_{4 4}&=& \frac{1}{3} \sin ^2\left(\theta _2\right) \sin ^2\left(\theta _3\right) \sin \left(2 \phi _2\right) \sin \left(2 \phi _3\right),\nonumber\\
t_{4 5}&=& \frac{1}{3} \sin ^2\left(\theta _2\right) \sin \left(2 \theta _3\right) \sin \left(2 \phi _2\right) \sin \left(\phi _3\right),\nonumber\\
t_{4 6}&=& \frac{\sin ^2\left(\theta _2\right) \left(3 \cos \left(2 \theta _3\right)+1\right) \sin \left(2 \phi _2\right)}{6 \sqrt{3}},\nonumber\\
t_{4 7}&=& \frac{1}{3} \sin ^2\left(\theta _2\right) \sin \left(2 \theta _3\right) \sin \left(2 \phi _2\right) \cos \left(\phi _3\right),\nonumber\\
t_{4 8}&=& \frac{1}{3} \sin ^2\left(\theta _2\right) \sin ^2\left(\theta _3\right) \sin \left(2 \phi _2\right) \cos \left(2 \phi _3\right),\nonumber\\
t_{50}&=& -\frac{\sin \left(2 \theta _2\right) \sin \left(\phi _2\right)}{3 \sqrt{3}},\nonumber\\
t_{05}&=& -\frac{\sin \left(2 \theta _3\right) \sin \left(\phi _3\right)}{3 \sqrt{3}},\nonumber\\
t_{5 4}&=& \frac{1}{3} \sin \left(2 \theta _2\right) \sin ^2\left(\theta _3\right) \sin \left(\phi _2\right) \sin \left(2 \phi _3\right),\nonumber\\
t_{5 5}&=& \frac{1}{3} \sin \left(2 \theta _2\right) \sin \left(2 \theta _3\right) \sin \left(\phi _2\right) \sin \left(\phi _3\right),\nonumber\\
t_{5 6}&=& \frac{\sin \left(2 \theta _2\right) \left(3 \cos \left(2 \theta _3\right)+1\right) \sin \left(\phi _2\right)}{6 \sqrt{3}},\nonumber\\
t_{5 7}&=& \frac{1}{3} \sin \left(2 \theta _2\right) \sin \left(2 \theta _3\right) \sin \left(\phi _2\right) \cos \left(\phi _3\right),\nonumber\\
t_{5 8}&=& \frac{1}{3} \sin \left(2 \theta _2\right) \sin ^2\left(\theta _3\right) \sin \left(\phi _2\right) \cos \left(2 \phi _3\right),\nonumber\\
t_{60}&=& \frac{1}{18} \left(-3 \cos \left(2 \theta _2\right)-1\right),\nonumber\\
t_{06}&=& \frac{1}{18} \left(-3 \cos \left(2 \theta _3\right)-1\right),\nonumber\\
t_{6 4}&=& \frac{\sin ^2\left(\theta _3\right) \left(3 \cos \left(2 \theta _2\right)+1\right) \sin \left(2 \phi _3\right)}{6 \sqrt{3}},\nonumber\\
t_{6 5}&=& \frac{\sin \left(2 \theta _3\right) \left(3 \cos \left(2 \theta _2\right)+1\right) \sin \left(\phi _3\right)}{6 \sqrt{3}},\nonumber\\
t_{6 6}&=& \frac{1}{36} \left(3 \cos \left(2 \theta _2\right)+1\right) \left(3 \cos \left(2 \theta _3\right)+1\right),\nonumber\\
t_{6 7}&=& \frac{\sin \left(2 \theta _3\right) \left(3 \cos \left(2 \theta _2\right)+1\right) \cos \left(\phi _3\right)}{6 \sqrt{3}},\nonumber\\
t_{6 8}&=& \frac{\sin ^2\left(\theta _3\right) \left(3 \cos \left(2 \theta _2\right)+1\right) \cos \left(2 \phi _3\right)}{6 \sqrt{3}},\nonumber\\
t_{70}&=& -\frac{\sin \left(2 \theta _2\right) \cos \left(\phi _2\right)}{3 \sqrt{3}},\nonumber\\
t_{07}&=& -\frac{\sin \left(2 \theta _3\right) \cos \left(\phi _3\right)}{3 \sqrt{3}},\nonumber\\
t_{7 4}&=& \frac{1}{3} \sin \left(2 \theta _2\right) \sin ^2\left(\theta _3\right) \sin \left(2 \phi _3\right) \cos \left(\phi _2\right),\nonumber
\end{eqnarray}

\begin{eqnarray}
t_{7 5}&=& \frac{1}{3} \sin \left(2 \theta _2\right) \sin \left(2 \theta _3\right) \sin \left(\phi _3\right) \cos \left(\phi _2\right),\nonumber\\
t_{7 6}&=& \frac{\sin \left(2 \theta _2\right) \left(3 \cos \left(2 \theta _3\right)+1\right) \cos \left(\phi _2\right)}{6 \sqrt{3}},\nonumber\\
t_{7 7}&=& \frac{1}{3} \sin \left(2 \theta _2\right) \sin \left(2 \theta _3\right) \cos \left(\phi _2\right) \cos \left(\phi _3\right),\nonumber\\
t_{7 8}&=& \frac{1}{3} \sin \left(2 \theta _2\right) \sin ^2\left(\theta _3\right) \cos \left(\phi _2\right) \cos \left(2 \phi _3\right),\nonumber\\
t_{80}&=& -\frac{\sin ^2\left(\theta _2\right) \cos \left(2 \phi _2\right)}{3 \sqrt{3}},\nonumber\\
t_{08}&=& -\frac{\sin ^2\left(\theta _3\right) \cos \left(2 \phi _3\right)}{3 \sqrt{3}},\nonumber\\
t_{8 4}&=& \frac{1}{3} \sin ^2\left(\theta _2\right) \sin ^2\left(\theta _3\right) \sin \left(2 \phi _3\right) \cos \left(2 \phi _2\right),\nonumber\\
t_{8 5}&=& \frac{1}{3} \sin ^2\left(\theta _2\right) \sin \left(2 \theta _3\right) \sin \left(\phi _3\right) \cos \left(2 \phi _2\right),\nonumber\\
t_{8 6}&=& \frac{\sin ^2\left(\theta _2\right) \left(3 \cos \left(2 \theta _3\right)+1\right) \cos \left(2 \phi _2\right)}{6 \sqrt{3}},\nonumber\\
t_{8 7}&=& \frac{1}{3} \sin ^2\left(\theta _2\right) \sin \left(2 \theta _3\right) \cos \left(2 \phi _2\right) \cos \left(\phi _3\right),\nonumber\\
t_{8 8}&=& \frac{1}{3} \sin ^2\left(\theta _2\right) \sin ^2\left(\theta _3\right) \cos \left(2 \phi _2\right) \cos \left(2 \phi _3\right).\nonumber
\end{eqnarray}
\section{Multipole parameters for $\chio\to\ff$}\label{appB}
We collected the multipole parameters for $\chio\to\ff$, i.e.,
\begin{eqnarray}
C_{00}&=&\frac{9}{8} \left(-\left(\cos \left(2 \theta _1\right)-5\right) \left|F^{(1)}_{0,1}\right|{}^2\right.\nonumber\\
&-&\left.\left(\cos \left(2 \theta _1\right)-5\right) \left|F^{(1)}_{1,0}\right|{}^2+2 \left(\cos \left(2 \theta _1\right)+3\right) \left|F^{(1)}_{1,1}\right|{}^2\right),\nonumber\\
C_{5 4}&=& \frac{3}{32} \sin \left(2 \theta _1\right) \left(F^{(1)\ast}_{1,1} F^{(1)}_{0,1}+F^{(1)\ast}_{0,1} F^{(1)}_{1,1}\right),\nonumber\\
C_{5 5}&=& \frac{3}{32} \left(\cos \left(2 \theta _1\right)-3\right) \left(F^{(1)\ast}_{1,0} F^{(1)}_{0,1}+F^{(1)\ast}_{0,1} F^{(1)}_{1,0}\right),\nonumber\\
C_{6 0}&=&\frac{1}{16} \left(2 \left(\cos \left(2 \theta _1\right)-5\right) \left|F^{(1)}_{0,1}\right|{}^2-\left(\cos \left(2 \theta _1\right)-5\right) \left|F^{(1)}_{1,0}\right|{}^2\right.\nonumber\\
   &+&2 \left.\left(\cos \left(2 \theta _1\right)+3\right) \left|F^{(1)}_{1,1}\right|{}^2\right),\nonumber\\
C_{0 6}&=&\frac{1}{16} \left(2 \left(\left(\cos \left(2 \theta _1\right)-5\right) \left|F^{(1)}_{1,0}\right|{}^2+\left(\cos \left(2 \theta _1\right)+3\right) \left|F^{(1)}_{1,1}\right|{}^2\right)\right.\nonumber\\
   &-&\left.\left(\cos \left(2 \theta _1\right)-5\right) \left|F^{(1)}_{0,1}\right|{}^2\right),\nonumber\\
C_{6 6}&=& \frac{1}{16} \left(\left(\cos \left(2 \theta _1\right)-5\right) \left|F^{(1)}_{0,1}\right|{}^2+\left(\cos \left(2 \theta _1\right)-5\right) \left|F^{(1)}_{1,0}\right|{}^2\right.\nonumber\\
&+&\left.\left(\cos \left(2 \theta _1\right)+3\right) \left|F^{(1)}_{1,1}\right|{}^2\right),\nonumber
\end{eqnarray}
\begin{eqnarray}
  C_{6 7}&=& \frac{1}{32} \sqrt{3} \sin \left(2 \theta _1\right) \left(F^{(1)\ast}_{1,1} F^{(1)}_{1,0}+F^{(1)\ast}_{1,0} F^{(1)}_{1,1}\right),\nonumber\\
  C_{6 8}&=& -\frac{1}{8} \sqrt{3} \sin ^2\left(\theta _1\right) \left|F^{(1)}_{0,1}\right|{}^2,\nonumber\\
  C_{7 0}&=&-\frac{1}{16} \sqrt{3} \sin \left(2 \theta _1\right) \left(F^{(1)\ast}_{1,1} F^{(1)}_{0,1}+F^{(1)\ast}_{0,1} F^{(1)}_{1,1}\right),\nonumber\\
  C_{0 7}&=&\frac{1}{16} \sqrt{3} \sin \left(2 \theta _1\right) \left(F^{(1)\ast}_{1,1} F^{(1)}_{1,0}+F^{(1)\ast}_{1,0} F^{(1)}_{1,1}\right),\nonumber\\
  C_{7 6}&=& -\frac{1}{32} \sqrt{3} \sin \left(2 \theta _1\right) \left(F^{(1)\ast}_{1,1} F^{(1)}_{0,1}+F^{(1)\ast}_{0,1} F^{(1)}_{1,1}\right),\nonumber\\
  C_{7 7}&=& \frac{3}{16} \left(F^{(1)\ast}_{1,0} F^{(1)}_{0,1}+F^{(1)\ast}_{0,1} F^{(1)}_{1,0}\right),\nonumber\\
  C_{7 8}&=& -\frac{3}{32} \sin \left(2 \theta _1\right) \left(F^{(1)\ast}_{1,1} F^{(1)}_{0,1}+F^{(1)\ast}_{0,1} F^{(1)}_{1,1}\right),\nonumber\\
  C_{8 0}&=&\frac{1}{8} \sqrt{3} \sin ^2\left(\theta _1\right) \left|F^{(1)}_{1,0}\right|{}^2,\nonumber\\
  C_{0 8}&=&\frac{1}{8} \sqrt{3} \sin ^2\left(\theta _1\right) \left|F^{(1)}_{0,1}\right|{}^2,\nonumber\\
  C_{8 6}&=& -\frac{1}{8} \sqrt{3} \sin ^2\left(\theta _1\right) \left|F^{(1)}_{1,0}\right|{}^2,\nonumber\\
  C_{8 7}&=& \frac{3}{32} \sin \left(2 \theta _1\right) \left(F^{(1)\ast}_{1,1} F^{(1)}_{1,0}+F^{(1)\ast}_{1,0} F^{(1)}_{1,1}\right),\nonumber\\
  C_{8 8}&=& -\frac{3}{16} \left(\cos \left(2 \theta _1\right)+3\right) \left|F^{(1)}_{1,1}\right|{}^2.\nonumber
\end{eqnarray}
\section{Multipole parameters for $\chit\to\ff$}\label{appC}
The multipole parameters for $\chit\to\ff$ are
\begin{eqnarray}
  C_{00}&=&\frac{9}{40} \left[-3 \cos \left(2 \theta _1\right)\right. \nonumber\\
  &\times&\left.\left(\left|F^{(2)}_{1,0}\right|{}^2+2 \left|F^{(2)}_{1,1}\right|{}^2+F^{(2)\ast}_{1,0} F^{(2)}_{0,1}-F^{(2)\ast}_{-1,1} F^{(2)}_{1,-1}\right)\right.\nonumber\\
&+&\left(7-3 \cos \left(2 \theta _1\right)\right) \left|F^{(2)}_{0,0}\right|{}^2+3 \left(\cos \left(2 \theta _1\right)+3\right) \left|F^{(2)}_{1,-1}\right|{}^2\nonumber\\
&+&15 \left|F^{(2)}_{1,0}\right|{}^2+14 \left|F^{(2)}_{1,1}\right|{}^2+\left.15 F^{(2)\ast}_{1,0} F^{(2)}_{0,1}+9 F^{(2)\ast}_{-1,1} F^{(2)}_{1,-1}\right],\nonumber\\
C_{4 4}&=& \frac{3}{80} \left(3 \cos \left(2 \theta _1\right)-7\right) \left|F^{(2)}_{1,1}\right|{}^2,\nonumber\\
C_{4 5}&=& -\frac{3}{160} \sqrt{3} \sin \left(2 \theta _1\right) \left(F^{(2)\ast}_{1,1} F^{(2)}_{1,0}+F^{(2)\ast}_{1,0} F^{(2)}_{1,1}\right),\nonumber\\
C_{5 4}&=& \frac{3}{160} \sqrt{3} \sin \left(2 \theta _1\right) \left(F^{(2)\ast}_{1,1} F^{(2)}_{1,0}+F^{(2)\ast}_{1,0} F^{(2)}_{1,1}\right)
,\nonumber\\
C_{5 5}&=& -\frac{3}{160} \left(6 \left(\cos \left(2 \theta _1\right)-3\right) \left|F^{(2)}_{1,0}\right|{}^2\right.\nonumber\\
      &+&\sqrt{6} \sin ^2\left(\theta _1\right) \left(F^{(2)\ast}_{1,-1} F^{(2)}_{0,0}+F^{(2)\ast}_{0,0} F^{(2)}_{1,-1}\right)\nonumber\\
&-&\left.\left(3 \cos \left(2 \theta _1\right)-7\right) \left(F^{(2)\ast}_{1,1} F^{(2)}_{0,0}+F^{(2)\ast}_{0,0} F^{(2)}_{1,1}\right)\right),\nonumber\\
\end{eqnarray}
\begin{eqnarray}
C_{6 0}&=&\frac{1}{40} \left(\left(3 \cos \left(2 \theta _1\right)-7\right) \left|F^{(2)}_{0,0}\right|{}^2+\left(7-3 \cos \left(2 \theta _1\right)\right) \left|F^{(2)}_{1,1}\right|{}^2\right.\nonumber\\
   &+&3\left. \left(\cos \left(2 \theta _1\right)+3\right) \left|F^{(2)}_{1,-1}\right|{}^2\right),\nonumber\\
C_{0 6}&=&\frac{1}{40} \left(\left(3 \cos \left(2 \theta _1\right)-7\right) \left|F^{(2)}_{0,0}\right|{}^2+\left(7-3 \cos \left(2 \theta _1\right)\right) \left|F^{(2)}_{1,1}\right|{}^2\right.\nonumber\\
 &+&3 \left.\left(\cos \left(2 \theta _1\right)+3\right) \left|F^{(2)}_{1,-1}\right|{}^2\right),\nonumber\\
\end{eqnarray}
\begin{eqnarray}
C_{6 6}&=& \frac{1}{80} \left(2 \left(7-3 \cos \left(2 \theta _1\right)\right) \left|F^{(2)}_{0,0}\right|{}^2+\left(7-3 \cos \left(2 \theta _1\right)\right) \left|F^{(2)}_{1,1}\right|{}^2\right.\nonumber\\
&+&\left.6 \left(\cos \left(2 \theta _1\right)-5\right) \left|F^{(2)}_{1,0}\right|{}^2+3 \left(\cos \left(2 \theta _1\right)+3\right) \left|F^{(2)}_{1,-1}\right|{}^2\right)
,\nonumber\\
C_{6 7}&=& -\frac{3}{160} \sin \left(2 \theta _1\right) \left(\left(2 F^{(2)\ast}_{0,0}-\sqrt{6} F^{(2)\ast}_{1,-1}+F^{(2)\ast}_{1,1}\right) F^{(2)}_{1,0}\right.\nonumber\\
  &+&F^{(2)\ast}_{1,0}\left. \left(2 F^{(2)}_{0,0}-\sqrt{6} F^{(2)}_{1,-1}+F^{(2)}_{1,1}\right)\right),\nonumber\\
C_{6 8}&=& \frac{3}{80} \sin ^2\left(\theta _1\right) \left(\sqrt{2} \left(F^{(2)\ast}_{1,1} F^{(2)}_{1,-1}+F^{(2)\ast}_{1,-1} F^{(2)}_{1,1}\right)-2 \sqrt{3} \left|F^{(2)}_{1,0}\right|{}^2\right),\nonumber\\
C_{7 6}&=& \frac{3}{160} \sin \left(2 \theta _1\right) \left(\left(2 F^{(2)\ast}_{0,0}-\sqrt{6} F^{(2)\ast}_{1,-1}+F^{(2)\ast}_{1,1}\right) F^{(2)}_{1,0}\right.\nonumber\\
&+&\left.F^{(2)\ast}_{1,0} \left(2 F^{(2)}_{0,0}-\sqrt{6} F^{(2)}_{1,-1}+F^{(2)}_{1,1}\right)\right),\nonumber\\
C_{7 7}&=& -\frac{3}{160} \left(12 \left|F^{(2)}_{1,0}\right|{}^2+\sqrt{6} \sin ^2\left(\theta _1\right) \left(F^{(2)\ast}_{1,-1} F^{(2)}_{0,0}+F^{(2)\ast}_{0,0} F^{(2)}_{1,-1}\right)\right.\nonumber\\
&+&\left.\left(3 \cos \left(2 \theta _1\right)-7\right) \left(F^{(2)\ast}_{1,1} F^{(2)}_{0,0}+F^{(2)\ast}_{0,0} F^{(2)}_{1,1}\right)\right)
,\nonumber\\
C_{7 8}&=& -\frac{3}{160} \sqrt{3} \sin \left(2 \theta _1\right) \left(F^{(2)\ast}_{1,1} F^{(2)}_{1,0}+F^{(2)\ast}_{1,0} F^{(2)}_{1,1}\right)
,\nonumber\\
C_{8 0}&=&\frac{3 \sin ^2\left(\theta _1\right) \left(F^{(2)\ast}_{1,1} F^{(2)}_{1,-1}+F^{(2)\ast}_{1,-1} F^{(2)}_{1,1}\right)}{20 \sqrt{2}}
,\nonumber\\
C_{0 8}&=&\frac{3 \sin ^2\left(\theta _1\right) \left(F^{(2)\ast}_{1,1} F^{(2)}_{1,-1}+F^{(2)\ast}_{1,-1} F^{(2)}_{1,1}\right)}{20 \sqrt{2}},\nonumber\\
C_{8 6}&=& \frac{3}{80} \sin ^2\left(\theta _1\right) \left(\sqrt{2} \left(F^{(2)\ast}_{1,1} F^{(2)}_{1,-1}+F^{(2)\ast}_{1,-1} F^{(2)}_{1,1}\right)-2 \sqrt{3} \left|F^{(2)}_{1,0}\right|{}^2\right)
,\nonumber\\
C_{8 7}&=& \frac{3}{160} \sqrt{3} \sin \left(2 \theta _1\right) \left(F^{(2)\ast}_{1,1} F^{(2)}_{1,0}+F^{(2)\ast}_{1,0} F^{(2)}_{1,1}\right),\nonumber\\
C_{8 8}&=& -\frac{3}{80} \left(3 \cos \left(2 \theta _1\right)-7\right) \left|F^{(2)}_{1,1}\right|{}^2.\nonumber
\end{eqnarray}

\section{Amplitudes of $\chi_{c1} \to \phi \phi$ decay}\label{app:chic1}
For $\chi_{c1} \to \phi \phi$ decay, the amplitudes corresponding to Fig. (\ref{fig:chic1-D-loop}) are
\begin{eqnarray}
\mathcal{M}_{(1-1)}&=&\int \frac{d^4 q}{(2\pi)^4} \frac{\tilde{g}_{\xi}^{\mu}(k_1)}{k_1^2 - m_{\mathcal{D}^\ast}^2} \frac{1}{k_2^2 - m_{\mathcal{D}}^2} \frac{1}{q^2 - m_{\mathcal{D}}^2} \mathcal{F}^2(q^2)\nonumber\\
&&\times[-ig_{\chi_{c1}\mathcal{D} \mathcal{D}^\ast}\epsilon^{\mu}_{\chi_{c1}}(p_1)] [-g_{\mathcal{D} \mathcal{D} \phi} \epsilon^{\ast\lambda}_{\phi}(p_3)(q_\lambda-k_{2\lambda})]\nonumber\\
&&\times[2f_{\mathcal{D} \mathcal{D}^\ast \phi} \varepsilon^{\zeta\eta\kappa\xi} \epsilon^{\ast}_{\phi\zeta}(p_2) p_{2\eta} (k_{1\kappa}+q_\kappa)],
\end{eqnarray}
\begin{eqnarray}
\mathcal{M}_{(1-2)}&=&\int \frac{d^4 q}{(2\pi)^4} \frac{1}{k_1^2 - m_{\mathcal{D}}^2} \frac{\tilde{g}_{\sigma\mu}(k_2)}{k_2^2 - m_{\mathcal{D}^\ast}^2} \frac{1}{q^2 - m_{\mathcal{D}}^2} \mathcal{F}^2(q^2)\nonumber\\
&&\times[ig_{\chi_{c1}\mathcal{D} \mathcal{D}^\ast}\epsilon^{\mu}_{\chi_{c1}}(p_1)] [-g_{\mathcal{D} \mathcal{D} \phi} \epsilon^{\ast\zeta}_{\phi}(p_2)(k_{1\zeta}+q_\zeta)]\nonumber\\
&&\times[-2f_{\mathcal{D} \mathcal{D}^\ast \phi} \varepsilon^{\lambda\rho\delta\sigma} \epsilon^{\ast}_{\phi\lambda}(p_3) p_{3\rho} (q_\delta-k_{2\delta})],
\end{eqnarray}
\begin{eqnarray}
\mathcal{M}_{(1-3)}&=&\int \frac{d^4 q}{(2\pi)^4} \frac{\tilde{g}^{\mu\psi}(k_1)}{k_1^2 - m_{\mathcal{D}^\ast}^2} \frac{1}{k_2^2 - m_{\mathcal{D}}^2} \frac{\tilde{g}_{\sigma}^{\gamma}(q)}{q^2 - m_{\mathcal{D}^\ast}^2} \mathcal{F}^2(q^2)\nonumber\\
&&\times[-ig_{\chi_{c1}\mathcal{D} \mathcal{D}^\ast}\epsilon^{\mu}_{\chi_{c1}}(p_1)] [g_{\mathcal{D}^\ast \mathcal{D}^\ast \psi} g_{\eta \gamma}g_{\psi}^{\eta} (k_{1\zeta} + q_\zeta)\nonumber\\
&&- 4f_{\mathcal{D}^\ast \mathcal{D}^\ast \phi} p_2^\eta (g_{\gamma\eta}g_{\psi\zeta} - g_{\gamma\zeta}g_{\psi\eta})]\epsilon^{\ast\zeta}_{\phi}(p_2) \nonumber\\
&&\times[2f_{\mathcal{D} \mathcal{D}^\ast \phi} \varepsilon^{\lambda\rho\delta\sigma} \epsilon^{\ast}_{\phi\lambda}(p_3) p_{3\rho} (q_\delta-k_{2\delta})],
\end{eqnarray}
\begin{eqnarray}
\mathcal{M}_{(1-4)}&=&\int \frac{d^4 q}{(2\pi)^4} \frac{1}{k_1^2 - m_{\mathcal{D}}^2} \frac{\tilde{g}^{\mu\iota}(k_2)}{k_2^2 - m_{\mathcal{D}^\ast}^2} \frac{\tilde{g}_{\xi}^{\upsilon}(q)}{q^2 - m_{\mathcal{D}^\ast}^2} \mathcal{F}^2(q^2)\nonumber\\
&&\times[-2f_{\mathcal{D} \mathcal{D}^\ast \phi} \varepsilon^{\zeta\eta\kappa\xi} \epsilon^{\ast}_{\phi\zeta}(p_2) p_{2\eta} (k_{1\kappa}+q_\kappa)]\nonumber\\
&&\times\epsilon^{\ast\lambda}_{\phi}(p_3) [g_{\mathcal{D}^\ast \mathcal{D}^\ast \phi} g_{\rho \iota}g_{\upsilon}^{\rho} (q_\lambda-k_{2\lambda})\nonumber\\
&&- 4f_{\mathcal{D}^\ast \mathcal{D}^\ast \phi} p_3^\rho (g_{\iota\rho}g_{\upsilon\lambda} - g_{\iota\lambda}g_{\upsilon\rho})]\nonumber\\
&&\times[ig_{\chi_{c1}\mathcal{D} \mathcal{D}^\ast}\epsilon^{\mu}_{\chi_{c1}}(p_1)],
\end{eqnarray}

\section{Amplitudes of $\chi_{c2} \to \phi \phi$ decay}\label{app:chic2}
The obtained amplitudes of the $\chi_{c2} \to \phi \phi$ decay for Fig. (\ref{fig:chic2-D-loop}) are
\begin{eqnarray}
\mathcal{M}_{(2-1)}&=&\int \frac{d^4 q}{(2\pi)^4} \frac{1}{k_1^2 - m_{\mathcal{D}}^2} \frac{1}{k_2^2 - m_{\mathcal{D}}^2} \frac{1}{q^2 - m_{\mathcal{D}}^2} \mathcal{F}^2(q^2)\nonumber\\
&&\times[g_{\chi_{c2}\mathcal{D}\mathcal{D}}\epsilon_{\chi_{c2}}^{\mu\nu}(p1) k_{2\mu} k_{1\nu}]\nonumber\\
&&\times [g_{\mathcal{D} \mathcal{D} \phi} \epsilon^{\ast\zeta}_{\phi}(p_2)(k_{1\zeta}+q_\zeta)]\nonumber\\
&&\times [g_{\mathcal{D} \mathcal{D} \phi} \epsilon^{\ast\lambda}_{\phi}(p_3)(q_\lambda-k_{2\lambda})],
\end{eqnarray}
\begin{eqnarray}
\mathcal{M}_{(2-2)}&=&\int \frac{d^4 q}{(2\pi)^4} \frac{\tilde{g}_{\xi}^{\tau}(k_1)}{k_1^2 - m_{\mathcal{D}^\ast}^2} \frac{1}{k_2^2 - m_{\mathcal{D}}^2} \frac{1}{q^2 - m_{\mathcal{D}}^2} \mathcal{F}^2(q^2)\nonumber\\
&&\times[-g_{\chi_{c2}\mathcal{D} \mathcal{D}^\ast} \varepsilon_{\mu\tau\alpha\beta} p_{1}^{\alpha} \epsilon^{\mu\nu}_{\chi_{c2}}(p_1) k_{2}^\beta k_{1\nu}]\nonumber\\
&&\times[2f_{\mathcal{D} \mathcal{D}^\ast \phi} \varepsilon^{\zeta\eta\kappa\xi} \epsilon^{\ast}_{\phi\zeta}(p_2) p_{2\eta} (k_{1\kappa}+q_\kappa)]\nonumber\\
&&\times [-g_{\mathcal{D} \mathcal{D} \phi} \epsilon^{\ast\lambda}_{\phi}(p_3)(q_\lambda-k_{2\lambda})],
\end{eqnarray}
\begin{eqnarray}
\mathcal{M}_{(2-3)}&=&\int \frac{d^4 q}{(2\pi)^4} \frac{1}{k_1^2 - m_{\mathcal{D}}^2} \frac{\tilde{g}_{\sigma}^{\tau}(k_2)}{k_2^2 - m_{\mathcal{D}^\ast}^2} \frac{1}{q^2 - m_{\mathcal{D}}^2} \mathcal{F}^2(q^2)\nonumber\\
&&\times[-g_{\chi_{c2}\mathcal{D} \mathcal{D}^\ast} \varepsilon_{\mu\tau\alpha\beta} p_{1}^{\alpha} \epsilon^{\mu\nu}_{\chi_{c2}}(p_1) k_{2\nu} k_{1}^{\beta}]\nonumber\\
&&\times [-g_{\mathcal{D} \mathcal{D} \phi} \epsilon^{\ast\zeta}_{\phi}(p_2)(k_{1\zeta}+q_\zeta)]\nonumber\\
&&\times[-2f_{\mathcal{D} \mathcal{D}^\ast \phi} \varepsilon^{\lambda\rho\delta\sigma} \epsilon^{\ast}_{\phi\lambda}(p_3) p_{3\rho} (q_\delta-k_{2\delta})],
\end{eqnarray}
\begin{eqnarray}
\mathcal{M}_{(2-4)}&=&\int \frac{d^4 q}{(2\pi)^4} \frac{1}{k_1^2 - m_{\mathcal{D}}^2} \frac{1}{k_2^2 - m_{\mathcal{D}}^2} \frac{\tilde{g}_{\xi\sigma}(q)}{q^2 - m_{\mathcal{D}^\ast}^2} \mathcal{F}^2(q^2)\nonumber\\
&&\times[g_{\chi_{c2}\mathcal{D}\mathcal{D}}\epsilon_{\chi_{c2}}^{\mu\nu}(p1) k_{2\mu} k_{1\nu}]\nonumber\\
&&\times[-2f_{\mathcal{D} \mathcal{D}^\ast \phi} \varepsilon^{\zeta\eta\kappa\xi} \epsilon^{\ast}_{\phi\zeta}(p_2) p_{2\eta} (k_{1\kappa}+q_\kappa)]\nonumber\\
&&\times[2f_{\mathcal{D} \mathcal{D}^\ast \phi} \varepsilon^{\lambda\rho\delta\sigma} \epsilon^{\ast}_{\phi\lambda}(p_3) p_{3\rho} (q_\delta-k_{2\delta})],
\end{eqnarray}
\begin{eqnarray}
\mathcal{M}_{(2-5)}&=&\int \frac{d^4 q}{(2\pi)^4} \frac{\tilde{g}_{\omega\xi}(k_1)}{k_1^2 - m_{\mathcal{D}^\ast}^2} \frac{\tilde{g}_{\chi\sigma}(k_2)}{k_2^2 - m_{\mathcal{D}^\ast}^2} \frac{1}{q^2 - m_{\mathcal{D}}^2} \mathcal{F}^2(q^2)\nonumber\\
&&\times[g_{\chi_{c2} \mathcal{D}^\ast \mathcal{D}^\ast} \epsilon^{\mu\nu}_{\chi_{c2}}(p_1) (g_{\nu\omega}g_{\mu\chi} + g_{\mu\omega}g_{\nu\chi})]\nonumber\\
&&\times[2f_{\mathcal{D} \mathcal{D}^\ast \phi} \varepsilon^{\zeta\eta\kappa\xi} \epsilon^{\ast}_{\phi\zeta}(p_2) p_{2\eta} (k_{1\kappa}+q_\kappa)]\nonumber\\
&&\times[-2f_{\mathcal{D} \mathcal{D}^\ast \phi} \varepsilon^{\lambda\rho\delta\sigma} \epsilon^{\ast}_{\phi\lambda}(p_3) p_{3\rho} (q_\delta-k_{2\delta})],
\end{eqnarray}
\begin{eqnarray}
\mathcal{M}_{(2-6)}&=&\int \frac{d^4 q}{(2\pi)^4} \frac{\tilde{g}^{\tau\psi}(k_1)}{k_1^2 - m_{\mathcal{D}^\ast}^2} \frac{1}{k_2^2 - m_{\mathcal{D}}^2} \frac{\tilde{g}_\sigma^\gamma(q)}{q^2 - m_{\mathcal{D}^\ast}^2} \mathcal{F}^2(q^2)\nonumber\\
&&\times[-g_{\chi_{c2}\mathcal{D} \mathcal{D}^\ast} \varepsilon_{\mu\tau\alpha\beta} p_{1}^{\alpha} \epsilon^{\mu\nu}_{\chi_{c2}}(p_1) k_{2}^\beta k_{1\nu}]\nonumber\\
&&\times\epsilon^{\ast\zeta}_{\phi}(p_2) [g_{\mathcal{D}^\ast \mathcal{D}^\ast \psi} g_{\eta \gamma}g_{\psi}^{\eta} (k_{1\zeta} + q_\zeta)\nonumber\\
&&- 4f_{\mathcal{D}^\ast \mathcal{D}^\ast \phi} p_2^\eta (g_{\gamma\eta}g_{\psi\zeta} - g_{\gamma\zeta}g_{\psi\eta})]\nonumber\\
&&\times[2f_{\mathcal{D} \mathcal{D}^\ast \phi} \varepsilon^{\lambda\rho\delta\sigma} \epsilon^{\ast}_{\phi\lambda}(p_3) p_{3\rho} (q_\delta-k_{2\delta})],
\end{eqnarray}
\begin{eqnarray}
\mathcal{M}_{(2-7)}&=&\int \frac{d^4 q}{(2\pi)^4} \frac{1}{k_1^2 - m_{\mathcal{D}}^2} \frac{\tilde{g}^{\tau\iota}(k_2)}{k_2^2 - m_{\mathcal{D}^\ast}^2} \frac{\tilde{g}_\xi^\upsilon(q)}{q^2 - m_{\mathcal{D}^\ast}^2} \mathcal{F}^2(q^2)\nonumber\\
&&\times[-g_{\chi_{c2}\mathcal{D} \mathcal{D}^\ast} \varepsilon_{\mu\tau\alpha\beta} p_{1}^{\alpha} \epsilon^{\mu\nu}_{\chi_{c2}}(p_1) k_{2\nu} k_{1}^{\beta}]\nonumber\\
&&\times[-2f_{\mathcal{D} \mathcal{D}^\ast \phi} \varepsilon^{\zeta\eta\kappa\xi} \epsilon^{\ast}_{\phi\zeta}(p_2) p_{2\eta} (k_{1\kappa}+q_\kappa)]\nonumber\\
&&\times \epsilon^{\ast\lambda}_{\phi}(p_3) [g_{\mathcal{D}^\ast \mathcal{D}^\ast \phi} g_{\rho \iota}g_{\upsilon}^{\rho} (q_\lambda-k_{2\lambda})\nonumber\\
&&- 4f_{\mathcal{D}^\ast \mathcal{D}^\ast \phi} p_3^\rho (g_{\iota\rho}g_{\upsilon\lambda} - g_{\iota\lambda}g_{\upsilon\rho})],
\end{eqnarray}
\begin{eqnarray}
\mathcal{M}_{(2-8)}&=&\int \frac{d^4 q}{(2\pi)^4} \frac{\tilde{g}_\omega^\psi(k_1)}{k_1^2 - m_{\mathcal{D}^\ast}^2} \frac{\tilde{g}_\chi^\iota(k_2)}{k_2^2 - m_{\mathcal{D}^\ast}^2} \frac{\tilde{g}^{\gamma\upsilon}(q)}{q^2 - m_{\mathcal{D}^\ast}^2} \mathcal{F}^2(q^2)\nonumber\\
&&\times[g_{\chi_{c2} \mathcal{D}^\ast \mathcal{D}^\ast} \epsilon^{\mu\nu}_{\chi_{c2}}(p_1) (g_{\nu\omega}g_{\mu\chi} + g_{\mu\omega}g_{\nu\chi})]\nonumber\\
&&\times \epsilon^{\ast\zeta}_{\phi}(p_2) [g_{\mathcal{D}^\ast \mathcal{D}^\ast \psi} g_{\eta \gamma}g_{\psi}^{\eta} (k_{1\zeta} + q_\zeta)\nonumber\\
&&- 4f_{\mathcal{D}^\ast \mathcal{D}^\ast \phi} p_2^\eta (g_{\gamma\eta}g_{\psi\zeta} - g_{\gamma\zeta}g_{\psi\eta})]\nonumber\\
&&\times \epsilon^{\ast\lambda}_{\phi}(p_3) [g_{\mathcal{D}^\ast \mathcal{D}^\ast \phi} g_{\rho \iota}g_{\upsilon}^{\rho} (q_\lambda-k_{2\lambda})\nonumber\\
&&- 4f_{\mathcal{D}^\ast \mathcal{D}^\ast \phi} p_3^\rho (g_{\iota\rho}g_{\upsilon\lambda} - g_{\iota\lambda}g_{\upsilon\rho})].
\end{eqnarray}

\end{appendices}

\vfil

\end{document}